\documentclass[conference]{./templates/IEEEtran}[journal]

\usepackage{graphicx}
\usepackage{booktabs}
\usepackage[table]{xcolor}

\usepackage{standalone}

\usepackage{tikz}
\usepackage{pgfplots}
\usepackage{pifont}  
\usepackage{indentfirst}
\usepackage{lipsum}
\usepackage{tabularx}
\usepackage{amsmath}
\usepackage{makecell}
\usetikzlibrary{spy}
\usepgfplotslibrary{groupplots}
\renewcommand{\arraystretch}{1.1} 
\newcommand{\cmark}{{\color{green!60!black}\ding{51}}}
\newcommand{\ccmark}{{\color{yellow!60!black}\ding{51}}}
\newcommand{\xmark}{{\color{red!70!black}\ding{55}}}

\newcommand{\pdftitle}{The Energy Cost of Artificial Intelligence Lifecycle in Communication Networks}

\usepackage[nolist,nohyperlinks]{acronym}
\usepackage{amsmath}
\usepackage{breqn}
\usepackage{balance}
\usepackage{cite}
\usepackage{framed}
\usepackage{siunitx}
\usepackage{soul}
\usepackage{multirow}
\usepackage{etoolbox}

\usepackage{blindtext}
\usepackage{booktabs}
\usepackage{footmisc} 
\usepackage{algorithm}
\usepackage{algpseudocode}
\usepackage{graphicx}
\usepackage{subcaption}

\usepackage{standalone}
\usepackage{tikz}
\usepackage{pgfplots}
\usetikzlibrary{spy}
\usepgfplotslibrary{groupplots}

\graphicspath{{./images/}}

\usepackage{url}
\usepackage[bookmarks=false]{hyperref}
\hypersetup{
 hidelinks,
 colorlinks=false,
 breaklinks=true,
 bookmarksopen=false,
 pdftitle=\pdftitle,
 pdfauthor={Author}%
}
\usepackage[noabbrev]{cleveref}

\newtoggle{authorcomment}
\toggletrue{authorcomment}

\begin{document}

\bstctlcite{IEEEexample:BSTcontrol}

\title{\pdftitle}


\author{
    Shih-Kai Chou\textsuperscript{1},
    Jernej Hribar\textsuperscript{1},
    Vid Hanžel\textsuperscript{1}, 
    Mihael Mohorčič\textsuperscript{1},
    Carolina Fortuna\textsuperscript{1}
}


\maketitle
\thispagestyle{plain}
\pagestyle{plain}

\begin{acronym}[MACHU]
  \acro{iot}[IoT]{Internet of Things}
  \acro{cr}[CR]{Cognitive Radio}
  \acro{ofdm}[OFDM]{orthogonal frequency-division multiplexing}
  \acro{ofdma}[OFDMA]{orthogonal frequency-division multiple access}
  \acro{scfdma}[SC-FDMA]{single carrier frequency division multiple access}
  \acro{rbi}[RBI]{ Research Brazil Ireland}
  \acro{rfic}[RFIC]{radio frequency integrated circuit}
  \acro{rf}[RF]{radio frequency}
  \acro{sdr}[SDR]{Software Defined Radio}
  \acro{sdn}[SDN]{Software Defined Networking}
  \acro{su}[SU]{Secondary User}
  \acro{ra}[RA]{Resource Allocation}
  \acro{qos}[QoS]{quality of service}
  \acro{usrp}[USRP]{Universal Software Radio Peripheral}
  \acro{mno}[MNO]{Mobile Network Operator}
  \acro{mnos}[MNOs]{Mobile Network Operators}
  \acro{gsm}[GSM]{Global System for Mobile communications}
   \acro{gsma}[GSMA]{Global System for Mobile communications Alliance}
   \acro{ran}[RAN]{radio access network}
  \acro{tdma}[TDMA]{Time-Division Multiple Access}
  \acro{fdma}[FDMA]{Frequency-Division Multiple Access}
  \acro{gprs}[GPRS]{General Packet Radio Service}
  \acro{msc}[MSC]{Mobile Switching Centre}
  \acro{bsc}[BSC]{Base Station Controller}
  \acro{umts}[UMTS]{universal mobile telecommunications system}
  \acro{Wcdma}[WCDMA]{Wide-band code division multiple access}
  \acro{wcdma}[WCDMA]{wide-band code division multiple access}
  \acro{cdma}[CDMA]{code division multiple access}
  \acro{lte}[LTE]{Long Term Evolution}
  \acro{papr}[PAPR]{peak-to-average power rating}
  \acro{hn}[HetNet]{heterogeneous networks}
  \acro{phy}[PHY]{physical layer}
  \acro{mac}[MAC]{medium access control}
  \acro{amc}[AMC]{adaptive modulation and coding}
  \acro{mimo}[MIMO]{multiple input multiple output}
  \acro{rats}[RATs]{radio access technologies}
  \acro{vni}[VNI]{visual networking index}
  \acro{rbs}[RB]{resource blocks}
  \acro{rb}[RB]{resource block}
  \acro{ue}[UE]{user equipment}
  \acro{cqi}[CQI]{Channel Quality Indicator}
  \acro{hd}[HD]{half-duplex}
  \acro{fd}[FD]{full-duplex}
  \acro{sic}[SIC]{self-interference cancellation}
  \acro{si}[SI]{self-interference}
  \acro{bs}[BS]{base station}
  \acro{fbmc}[FBMC]{Filter Bank Multi-Carrier}
  \acro{ufmc}[UFMC]{Universal Filtered Multi-Carrier}
  \acro{scm}[SCM]{Single Carrier Modulation}
  \acro{isi}[ISI]{inter-symbol interference}
  \acro{ftn}[FTN]{Faster-Than-Nyquist}
  \acro{m2m}[M2M]{machine-to-machine}
  \acro{mtc}[MTC]{machine type communication}
  \acro{mmw}[mmWave]{millimeter wave}
  \acro{bf}[BF]{beamforming}
  \acro{los}[LOS]{line-of-sight}
  \acro{nlos}[NLOS]{non line-of-sight}
  \acro{capex}[CAPEX]{capital expenditure}
  \acro{opex}[OPEX]{operational expenditure}
  \acro{ict}[ICT]{information and communications technology}
  \acro{sp}[SP]{service providers}
  \acro{inp}[InP]{infrastructure providers}
  \acro{mvnp}[MVNP]{mobile virtual network provider}
  \acro{mvno}[MVNO]{mobile virtual network operator}
  \acro{nfv}[NFV]{network function virtualization}
  \acro{vnfs}[VNF]{virtual network functions}
  \acro{cran}[C-RAN]{Cloud Radio Access Network}
  \acro{vran}[V-RAN]{Virtual Radio Access Network}
  \acro{bbu}[BBU]{baseband unit}
  \acro{bbus}[BBU]{baseband units}
  \acro{rrh}[RRH]{remote radio head}
  \acro{rrhs}[RRH]{Remote radio heads} 
  \acro{sfv}[SFV]{sensor function virtualization}
  \acro{wsn}[WSN]{wireless sensor networks} 
  \acro{bio}[BIO]{Bristol is open}
  \acro{vitro}[VITRO]{Virtualized dIstributed plaTfoRms of smart Objects}
  \acro{os}[OS]{operating system}
  \acro{www}[WWW]{world wide web}
  \acro{iotvn}[IoT-VN]{IoT virtual network}
  \acro{mems}[MEMS]{micro electro mechanical system}
  \acro{mec}[MEC]{Mobile edge computing}
  \acro{coap}[CoAP]{Constrained Application Protocol}
  \acro{vsn}[VSN]{Virtual sensor network}
  \acro{rest}[REST]{REpresentational State Transfer}
  \acro{aoi}[AoI]{Age of Information}
  \acro{lora}[LoRa\texttrademark]{Long Range}
  \acro{iot}[IoT]{Internet of Things}
  \acro{snr}[SNR]{Signal-to-Noise Ratio}
  \acro{cps}[CPS]{Cyber-Physical System}
  \acro{uav}[UAV]{Unmanned Aerial Vehicle}
  \acro{rfid}[RFID]{Radio-frequency identification}
  \acro{lpwan}[LPWAN]{Low-Power Wide-Area Network}
  \acro{lgfs}[LGFS]{Last Generated First Served}
  \acro{wsn}[WSN]{wireless sensor network} 
  \acro{lmmse}[LMMSE]{Linear Minimum Mean Square Error}
  \acro{rl}[RL]{Reinforcement Learning}
  \acro{nb-iot}[NB-IoT]{Narrowband IoT}
  \acro{lorawan}[LoRaWAN]{Long Range Wide Area Network}
  \acro{mdp}[MDP]{Markov Decision Process}
  \acro{ann}[ANN]{Artificial Neural Network}
  \acro{dqn}[DQN]{Deep Q-Network}
  \acro{mse}[MSE]{Mean Square Error}
  \acro{ml}[ML]{Machine Learning}
  \acro{cpu}[CPU]{Central Processing Unit}
  \acro{gpu}[GPU]{Graphics Processing Unit}
  \acro{tpu}[TPU]{Tensor Processing Unit}
  \acro{ddpg}[DDPG]{Deep Deterministic Policy Gradient}
  \acro{ai}[AI]{Artificial Intelligence}
  \acro{gp}[GP]{Gaussian Processes}
  \acro{drl}[DRL]{Deep Reinforcement Learning}
  \acro{mmse}[MMSE]{Minimum Mean Square Error}
  \acro{fnn}[FNN]{Feedforward Neural Network}
  \acro{eh}[EH]{Energy Harvesting}
  \acro{wpt}[WPT]{Wireless Power Transfer}
  \acro{dl}[DL]{Deep Learning}
  \acro{yolo}[YOLO]{You Only Look Once}
  \acro{mec}[MEC]{Mobile Edge Computing}
  \acro{marl}[MARL]{Multi-Agent Reinforcement Learning}
  \acro{aiot}[AIoT]{Artificial Intelligence of Things}
  \acro{ru}[RU]{Radio Unit}
  \acro{du}[DU]{Distributed Unit}
  \acro{cu}[CU]{Central Unit}
  \acro{api}[API]{Application Programming Interface}
  \acro{cf}[CF]{Carbon Footprint}
  \acro{ci}[CI]{Carbon Intensity}
  \acro{oran}[O-RAN]{Open Radio Access Network}
  \acro{flops}[FLOPs]{floating-point operations per cycle per core}
  \acro{FLOPS}[FLOPs]{floating-point operations}
  \acro{5g}[5G]{fifth-generation}
  \acro{embb}[eMBB]{enhanced Mobile Broadband}
  \acro{urllc}[URLLC]{Ultra Reliable Low Latency Communications}
  \acro{mmtc}[mMTC]{massive Machine Type Communications}
  \acro{hdd}[HDD]{hard disk drive}
  \acro{ssd}[SSD]{solid state drive}
  \acro{fspl}[FSPL]{free space path loss}
  \acro{cav}[CAV]{Connected Autonomous Vehicle}
  \acro{xr}[XR]{Extended Reality}
  \acro{ap}[AP]{access point}
  \acro{ble}[BLE]{Bluetooth low energy}
  \acro{uwb}[UWB]{ultra wide band}
  \acro{rss}[RSS]{received signal strength}
  \acro{nn}[NN]{Neural Network}
  \acro{v2x}[V2X]{vehicle-to-everything}
  \acro{gai}[GAI]{Generative AI}
  \acro{aiaas}[AIaaS]{AI as a service}
  \acro{mlp}[MLP]{Multilayer Perceptron}
  \acro{ble}[BLE]{Bluetooth Low Energy}
  \acro{lorawan}[LoRaWAN]{Long Range Wide Area Network}
  \acro{FPGAs}[FPGAs]{Field Programmable Gate Arrays}
  \acro{llm}[LLM]{large language model}
  \acro{fl}[FL]{federated learning}
  \acro{pue}[PUE]{Power Usage Effectiveness}
  \acro{cue}[CUE]{Carbon Usage Effectiveness}
  \acro{wue}[WUE]{Water Usage Effectiveness}
  \acro{ee}[EE]{energy efficiency}
  \acro{kpi}[KPI]{Key Performance Index}
  \acro{dp}[DP]{Data Plane}
  \acro{cp}[CP]{Control Plane}
  \acro{aes}[AES]{Advanced Encryption Standard}
  \acro{osi}[OSI]{Open Systems Interconnection}
  \acro{cnn}[CNN]{Convolutional Neural Network}
  \acro{kan}[KAN]{Kolmogorov–Arnold Network}
  \acro{gadf}[GADF]{Gramian Angular Difference Field}
  \acro{PUE}[PUE]{Power Usage Effectiveness}
  \acro{CUE}[CUE]{Carbon Usage Effectiveness}
  \acro{WUE}[WUE]{Water Usage Effectiveness}
  \acro{aes}[AES]{Advanced Encryption Standard}
  \acro{phy}[PHY]{physical layer}
  \acro{nic}[NIC]{Network Interface Card}
  \acro{dma}[DMA]{Direct Memory Access}
  \acro{pc}[PC]{Personal Computer}
\end{acronym}

\begin{abstract}
Artificial Intelligence (AI) is being incorporated in several optimization, scheduling, orchestration as well as in native communication network functions. This paradigm shift results in increased energy consumption, however, quantifying the end-to-end energy consumption of adding intelligence to communication systems remains an open challenge since conventional energy consumption metrics focus on either communication, computation infrastructure, or model development. To address this, we propose a new metric, the Energy Cost of AI Lifecycle (eCAL) of an AI model in a system. eCAL captures the energy consumption throughout the development, deployment and utilization of an AI-model providing intelligence in a communication network by (i) analyzing the complexity of data collection and manipulation in individual components and (ii) deriving overall and per-bit energy consumption. We show that as a trained AI model is used more frequently for inference, its energy cost per inference decreases, since the fixed training energy is amortized over a growing number of inferences. 
For a simple case study we show that eCAL for $100$ inferences is $2.73$ times higher than for $1000$ inferences. Additionally, we have developed a modular and extendable open-source simulation tool to enable researchers, practitioners, and engineers to calculate the end-to-end energy cost with various configurations and across various systems, ensuring adaptability to diverse use cases.%

\end{abstract}
\renewcommand{\thefootnote}{\arabic{footnote}}
\footnotetext[1]{Jožef Stefan Institute, Ljubljana, 1000, Slovenia. Corresponding author: Shih-Kai Chou (e-mail: shih-kai.chou@ijs.si)}
\acresetall


\section{Introduction}
\label{sec:intro}
Next-generation wireless networks aim to deliver unprecedented levels of connectivity, service diversity and flexibility. \ac{ai} is expected to become fundamental to 6G and future cellular networks~\cite{letaief2019roadmap}, with traditional networking functions increasingly being supplanted by \ac{ai}‑driven implementations to realize so‑called “\ac{ai}‑native” capabilities~\cite{hoydis2021toward,wu2022ai}. These \ac{ai} techniques aim to deliver intelligent, sustainable, and dynamically programmable services that enhance network adaptability and flexibility~\cite{letaief2019roadmap}. Recent works propose a pervasive multi-level native \ac{ai} architecture that incorporates knowledge graph (KG) into mobile network reducing data scale and computational costs of \ac{ai} training by almost an order of magnitude \cite{10553365}, attempt to improve communication network efficiency by minimizing the transmitted data through semantic communications \cite{10554663}, and optimize end-to-end network slicing \cite{10636776}. Furthermore, and \ac{ai}/\ac{ml} workflow was defined ~\cite{b1} as part of the \ac{oran} Alliance that tackles the standardization of the network access segment.

While such systems designed through endogenous \ac{ai} models enable unprecedented automation and agility as well as improved energy consumption for the performed data transmission, their reliance on \ac{ai} implies additional energy consumption and increased carbon emissions~\cite{dhar2020carbon}. These energy and carbon costs may be significant \cite{luccioni2023estimating}, especially in the cases where models with many parameters are employed, such as \acp{llm}. Projections indicate that by 2030, electricity consumption could rise to 100 TWh for data centers and 40 TWh for telecommunication networks \cite{kamiya2024energy}. 

In light of climate challenges, the general \ac{ai} research community's efforts are intensifying. They aim to better estimate the costs of training~\cite{GARCIAMARTIN201975} and using \ac{ai}~\cite{luccioni2020estimating}, assess the \ac{cf} of \ac{llm}s~\cite{faiz2024llmcarbon}, and find ways to scale models efficiently~\cite{wu2024beyond}. Furthermore, while increasingly relying on \ac{ai} techniques, the networking community is also exploring pathways towards net-zero carbon emissions~\cite{AIoT_CE_AI}. In particular, researchers analyze the \ac{cf} of different learning techniques \cite{AIoT_CE_AI2} and data modalities and develop different approaches to optimize energy efficiency, including joint optimization of hardware-software co-design~\cite{aiot_ec1}, improved scheduling~\cite{aiot_ec2}, more efficient model design \cite{bertalanivc2022resource}, and energy/carbon consumption testing \cite{AIoT_CE_AI4}.

Depending on the scope of the study, various metrics are employed to understand \ac{ee} and \ac{cf}. As well noted in \cite{AIoT_CE_AI}, the traditional ITU standardized Energy-per-Bit $[J/b]$ metric \textit{'will no longer be able to reflect the environmental impact of the modern mobile services, especially network \ac{ai}-enabled smart services'}. For instance, \cite{aiot_ec1} considers both normalized energy and energy for buffering (in $[J]$), whereas \cite{AIoT_CE_AI,AIoT_CE_AI2} examine electricity consumption (in $[Wh]$) and carbon emissions (in $[CO_{2}eq]$) for the various phases involved in \ac{fl} edge systems. The energy cost and computational complexity  ($[J]$ and GFLOPS) of neural architectures are evaluated in \cite{bertalanivc2022resource}, while~\cite{aiot_ec2} focuses on energy cost $[Wh]$ of scheduled loads. However, \textit{these metrics are not specifically designed to capture the \ac{ee} of a system or its parts}. To this end, a metric somewhat similar to the Energy-per-Bit $[J/b]$ would be more suitable for measuring \ac{ee} of an \ac{ai}-based communication system, similarly as some other efficiency metrics that have been proposed in wireless systems in particular and in engineering and economy in general \cite{naastepad2003labour}. For instance, spectral efficiency, measured in $[b/s/Hz]$ \cite{rysavy2014challenges} provides means of calculating the amount of data bandwidth available in a given amount of spectrum, and lifecycle emissions for vehicles, measured in $[g/km]$ \cite{BUBERGER2022112158}, enable computing the \ac{cf} in grams per kilometer.

Following these observations, we tackle a timely, unexplored and complex problem aimed to find a suitable \ac{ee} metric that would be simple and general and would holistically quantify the environmental cost of inference carried out in future communication networks. This work is meant as the foundation on which to build increasingly accurate understanding of the cost of AI model embedding in networks. The contributions of this paper are as follows:
\begin{itemize}
    \item We propose a new metric, the Energy Cost of AI Lifecycle (eCAL), measured in $[J/b]$,  that captures the overall energy cost of generating an inference in a communication system. Unlike capturing the energy required for transmitting bits that is done by the Energy-per-Bit $[J/b]$ metric, or $[J]$ for \ac{ai} model complexity, the proposed eCAL captures the energy consumed by all the data collection and manipulation components in an AI-enhanced system during the lifecycle of a trained model delivering inference. 
    \item Following the standard \ac{osi} model and AI/ML workflow or Machine Learning Operations (MLOps) \cite{diaz2023joint}, falling under the platform layer of the standard Cloud Computing Reference Architecture (CCRA), we devise a methodology for determining the eCAL of an communication system and develop an open source modular and extensible eCAL calculator\footnote{https://github.com/sensorlab/eCAL}. The methodology breaks down the system into different data manipulation components,  and, for each component, it analyzes the complexity of data manipulation and derives the overall energy consumption. 
    \item Using the proposed metric, we demonstrate that the better a model and the more it is used, the more energy-efficient is inference. For a selected case study, the energy consumption per bit for $100$ inferences is $2.73$ times higher than for $1000$ inferences. This can be explained by the fact that the energy cost of data collection, preprocessing, training, and evaluation, all necessary to develop an \ac{ai} model, are spread across more inferences.    
\end{itemize}

The paper is organized as follows. Section~\ref{sec:related} reviews related work. Section~\ref{sec:method} describes the AI model lifecycle, introduces eCAL, and outlines its derivation methodology. Sections~\ref{sec:E_T}–\ref{sec:E_inf} present energy consumption formulas for the data manipulation components of the lifecycle. Section~\ref{sec:e2e} derives and analyzes eCAL for selected communication and model development cases, illustrating its effectiveness in capturing overall lifecycle energy costs. Section~\ref{sec:conclusion} concludes the paper and discusses future research directions.

\section{Related Work}
\label{sec:related}

\paragraph*{AI Environmental Impact} The growing dependence on \ac{ai} has introduced significant energy and carbon costs \cite{dhar2020carbon}, intensifying environmental concerns~\cite{faiz2024llmcarbon}. In response, the \ac{ai} research community has placed greater emphasis on understanding and then formalizing methods to assess the environmental impact of the increasing \ac{ai} adoption. Due to the complexity of state-of-the-art neural network architectures, in many cases the energy cost of training is computed after the training is done, by measuring the performed computation through interfaces such as the performance application programming interface \cite{barry2021effortless}. Furthermore, \cite{GARCIAMARTIN201975} showed that \textit{the energy footprint of inference has been more extensively studied than that of training}. The study highlighted that the accuracy between the predicted and measured energy consumption is below 70\%. 

\paragraph*{Emerging in the AI Community} A number of approaches and tools capable of proactively estimating the energy and environmental cost of training have also emerged. LLMCarbon \cite{faiz2024llmcarbon} is a very recent end-to-end \ac{cf} projection model designed for both dense (all parameters are used for every input) and mixture of experts (only a subset of parameters for each input) LLMs.  It incorporates critical LLM, hardware, and data center parameters, such as LLM parameter count, hardware type, system power, chip area, and data center efficiency, to model both operational and embodied \ac{cf}s of an LLM. Furthermore, \cite{wu2024beyond} aimed to provide guidelines for scaling \ac{ai} in a sustainable way. They examined the \ac{ee} of new processing units and the \ac{cf} of the most prominent LLM models since GPT-3, analyzed their lifecycle carbon impact, and showed that inference and training are comparable. They concluded that to enable sustainability as a computer system design principle, better tools for carbon telemetry, large-scale carbon datasets, carbon impact disclosure, and more suitable carbon metrics are required.

\begin{table*}[t]
\caption{State-of-the-art energy efficiency metrics, their relevance across the AI lifecycle, and additional system context.}
\label{tab:3gpp}
\centering
\small
\setlength{\tabcolsep}{3.5pt}
\renewcommand{\arraystretch}{1.15}
\resizebox{1\textwidth}{!}{%
\begin{tabular}{|l|
*{10}{>{\centering\arraybackslash}m{0.08\textwidth}}
|}
\hline
\textbf{Column} & \textbf{(2)} & \textbf{(3)} & \textbf{(4)} & \textbf{(5)} & \textbf{(6)} & \textbf{(7)} & \textbf{(8)} & \textbf{(9)} & \textbf{(10)} & \textbf{(11)} \\
\hline
\multirow{2}{*}{\textbf{Characteristic}} &
\multicolumn{10}{c|}{\textbf{Metrics}} \\
\cline{2-11}
& \makecell{\textbf{Energy}\\\textbf{Efficiency}} & \makecell{\textbf{Energy}\\\textbf{per Bit}} & \textbf{PUE} & \textbf{CUE} & \textbf{WUE} & \textbf{APC} & \textbf{APEC} & \textbf{TTCAPC} & \textbf{TTCAPEC} & \textbf{eCAL} \\
\hline
\textbf{Category} & Telecom & Telecom & Data Center & Data Center & Data Center & Deep Learning & Deep Learning & Deep Learning & Deep Learning & Proposed \\
\textbf{Units} & $b/J$ & $J/b$ & \% & $kg/kWh$ & $L/kWh$ & \% & \% & \% & \% & $J/b$ \\
\textbf{Description} & Measures information bits per Joule consumed by the radio access network or communication module. & Energy consumed to transmit one bit over the link. & Power Usage Effectiveness: ratio of total facility energy to energy delivered to IT. & Carbon Usage Effectiveness: CO$_2$ ($kg$) per kWh consumed. & Water Usage Effectiveness: liters of water per kWh for cooling/operation. & Accuracy per Consumption: accuracy relative to the energy cost of model weights (GreenAI). & Accuracy per Energy Cost: variant of APC capturing broader costs. & Time to Closest APC: integrates training time and energy into APC. & Time to Closest APEC: integrates training time and energy into APEC. & End-to-end energy of AI data collection/ inference relative to application bits (comms + compute). \\
\hline
\rowcolor{blue!10}
\multicolumn{11}{|c|}{\textbf{Mapping to the telco OSI standard layers and cloud computing CCRA standard layers.}} \\
\rowcolor{blue!10}
\hline
\textbf{OSI Standard} & Physical & Physical & All & All & All & Application & Application & Application & Application & All \\
\rowcolor{blue!10}
\textbf{CCRA Standard} & \xmark & \xmark & All & All & All & Service & Service & Service & Service & Service, Platform \\
\hline
\rowcolor{blue!10}\multicolumn{11}{|c|}{\textbf{Relevance to the data-manipulating components in the AI lifecycle}} \\
\hline
\rowcolor{blue!10}
\textbf{Data Collection (Sec.~IV)} & \cmark & \cmark & \ccmark & \ccmark & \ccmark & \xmark & \xmark & \xmark & \xmark & \cmark \\
\rowcolor{blue!10}
\textbf{Preproc. (Sec.~V)} & \xmark & \xmark & \ccmark & \ccmark & \ccmark & \xmark & \xmark & \xmark & \xmark & \cmark \\
\rowcolor{blue!10}
\textbf{Training (Sec.~VI)} & \xmark & \xmark & \ccmark & \ccmark & \ccmark & \cmark & \cmark & \cmark & \cmark & \cmark \\
\rowcolor{blue!10}
\textbf{Evaluation (Sec.~VII)} & \xmark & \xmark & \ccmark & \ccmark & \ccmark & \cmark & \cmark & \cmark & \cmark & \cmark \\
\rowcolor{blue!10}
\textbf{Inference (Sec.~VIII)} & \xmark & \xmark & \ccmark & \ccmark & \ccmark & \cmark & \cmark & \xmark & \xmark & \cmark \\
\hline
\end{tabular}%
}
\end{table*}

\paragraph*{Emerging in the Networking Community} Increasingly relying on \ac{ai} techniques, the networking community is also investigating pathways towards net-zero carbon emissions~\cite{AIoT_CE_AI}. In~\cite{AIoT_CE_AI} the authors noticed that in spite of improvements in hardware and software energy efficiency, the overall energy consumption of mobile networks continues to rise, exacerbated by the growing use of resource-intensive AI algorithms. They introduced a novel evaluation framework to analyze the lifecycle of network AI implementations, identifying major carbon emission sources. They proposed the dynamic energy trading and task allocation framework, designed to optimize carbon emissions by reallocating renewable energy sources and distributing tasks more efficiently across the network. Similar to \cite{wu2024beyond}, they highlighted the need for the development of new metrics to quantify the environmental impact of new network services enabled by AI.  

The authors of \cite{AIoT_CE_AI2} introduced a novel framework to quantify energy consumption and carbon emissions for vanilla \ac{fl} methods and consensus-based decentralized approaches, and identified optimal operational points for sustainable \ac{fl} designs. Two case studies were analyzed within 5G industry verticals: continual learning and reinforcement learning scenarios. Similar to the authors of \cite{AIoT_CE_AI}, they considered energy [$Wh$] and carbon emissions [$CO_{2}eq$] in their evaluations. The solution proposed in \cite{aiot_ec1} is a hardware/software co-design that introduces modality gating (throttling) to adaptively manage sensing and computing tasks. It features a novel decoupled modality sensor architecture that supports partial throttling of sensors, significantly reducing energy consumption while maintaining data flow across multimodal data: text, speech, images, and video. More energy efficient task scheduling has been considered in ~\cite{aiot_ec2}, while more efficient neural network architecture design and subsequent model development in \cite{bertalanivc2022resource}. \ac{ee} of programming languages was studied in \cite{pereira2021ranking}, while energy and carbon consumption testing for AI-driven \ac{iot} services were developed in \cite{AIoT_CE_AI4}.
\paragraph*{Existing Metrics and Scope Mapping to OSI and CCRA standards} Table~\ref{tab:3gpp} summarizes the relevant existing telecommunications (columns 2 and 3), data center (columns 4-6), and deep learning (columns 7-10) metrics proposed to quantify the energy cost of data collection and manipulation. The telecommunication metrics are well established, however they measure \ac{ee} and energy per bit respectively on the physical layer as summarized in the table. These \ac{ee} metrics were sufficient for previous generations of mobile communication systems, but they cannot capture the AI aspects to be used for implementing 6G network functionality as well as for optimizations \cite{AIoT_CE_AI}. Furthermore, according to~\cite{startingpoint}, standardization organizations have recently placed a greater emphasis on \ac{ee} and energy savings as a core topic, and 3GPP, for instance, has listed \ac{ee} as \ac{kpi} in its Release 19\footnote{https://www.3gpp.org/specifications-technologies/releases/release-19} set of specifications.  

The data center consumption metrics in Table \ref{tab:3gpp}, i.e. \ac{PUE}, \ac{CUE}, and \ac{WUE} \cite{datacenter_metric}, refer to the overall data center operation and include computing for AI and non AI purposes, networking and cooling. They offer a macro-system view agnostic to specifics of AI or communications as can be seen from their description in the table (see row "description"). They implicitly quantify the energy consumption across all layers of the \ac{osi} and CCRA as shown in the rows labeled OSI Standard and CCRA Standard, but they are not designed to provide per layer analytical insights and the energy per amount of processed information.

The \ac{dl} metrics summarized in columns 7-10, namely Accuracy Per Consumption (APC), Accuracy Per Energy Cost (APEC) for inference, Time To Closest APC (TTCAPC) and Time To Closest APEC (TTCAPEC) \cite{jurj2020environmentally}, 
consider not only the accuracy and speed of \ac{dl} models but also their energy consumption and cost for training. Nevertheless, while powerful in terms of assessing the energy/performance trade-offs of models, these metrics are less suitable in assessing the energy costs of adding intelligence to future communication networks, especially in cases when the dominant energy consumption arises from data collection rather than from the model training/inference. As marked in the table (OSI Standard and CCRA Standard), their scope falls under the application layer in \ac{osi} and service layer in CCRA. 

\paragraph*{Summary} Inspired by the simplicity and effectiveness of metrics such as transmission efficiency through energy-per-bit and the lifecycle emissions, and responding to calls for new metrics \cite{AIoT_CE_AI,wu2024beyond} that capture the energy cost in the era of \ac{ai} enabled systems, this work proposes the novel Energy Cost of AI Lifecycle (eCAL) metric and methodology to derive it. eCAL is the only metric that, based on \ac{osi} and CCRA, provides insights into the cost of inference, considering the data transmission and AI model overheads.

\section{Definition and Methodology}
\label{sec:method}
\begin{figure}[t]
    \centering
    \includegraphics[width=0.5\textwidth]{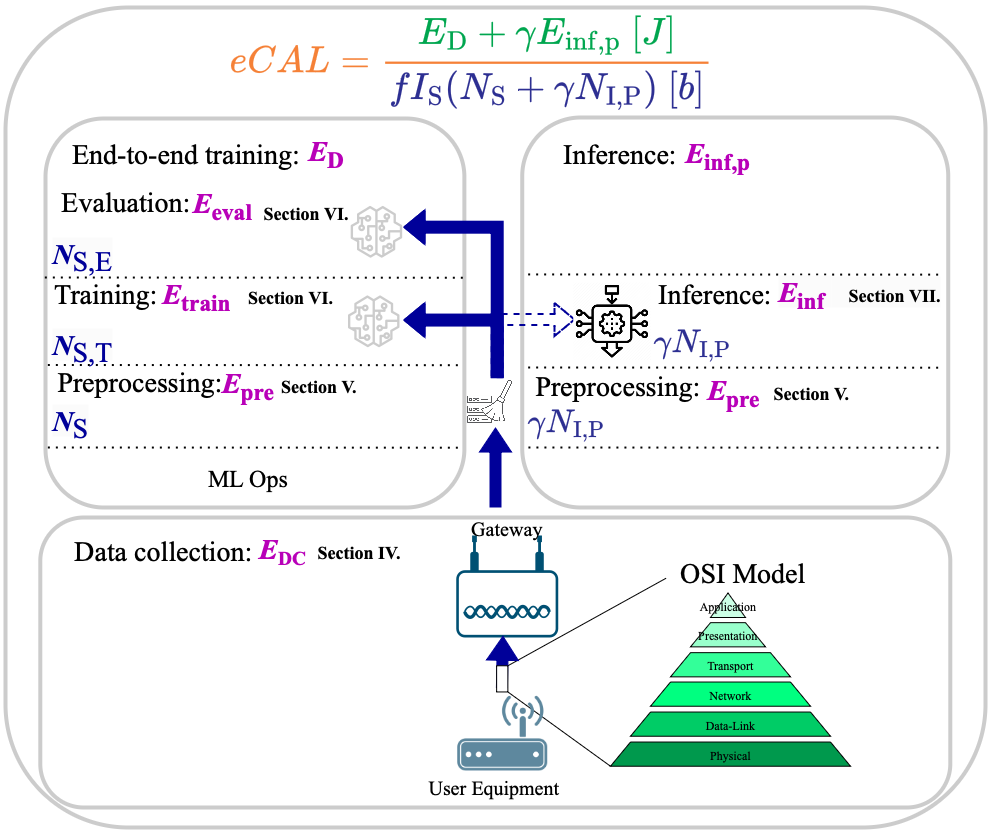}
    \caption{Data manipulation components involved in the lifecycle of an AI model.}
    \label{fig:system_model}
\end{figure}

In our work, we consider a communication and computing system that enables the development of an AI model, whose \textit{lifecycle} is depicted in Fig.~\ref{fig:system_model}, consisting of the following data manipulating components as per \ac{osi} and MLOps: 

\paragraph{Data Collection}\label{par:a} 
This data-manipulating component, depicted in the lower part of Fig.~\ref{fig:system_model}, represents receiving $N_{\mathrm{S}}$ data samples with $I_{\mathrm{S}}$ sample size from distributed devices via $N_{\mathrm{L}}$ independent wired or wireless links at the target computing infrastructure. The collected information can be turned into indicators such as signal strength and be interpreted by the \ac{ai} models, for example, to predict the location of a user. To successfully collect all the application-level data, the total energy of data collection $E_{\mathrm{DC}}~[J]$ is needed. The number, topology or other particularities of the data collection component of the AI model lifecycle can be adapted to specific configurations if needed.

\paragraph{Data Preprocessing}
\label{par:b} 
To ensure accuracy and reliability during the training process, the data must go through several preprocessing steps, such as cleaning, feature engineering and transformation. The energy consumption for preprocessing, $E_{\mathrm{pre}}$, depends on the integrity of the ingested dataset. The proposed metric is conceptually capable of supporting also distributed or hybrid pre-processing as well as training and evaluation scenarios. 

\paragraph{Training and Evaluation}\label{par:c} 
The training component comprises the \ac{ai}/\ac{ml} model development using selected \ac{ai}/\ac{ml} techniques, such as neural network architectures, and the training data. In this step of the model development process, the processed data, $N_{\mathrm{S,T}}$ is fetched and utilized to learn weights and biases in order to approximate the underlying distribution. For most neural network architectures, the learning processes rely on \acp{gpu} and \acp{tpu}, which perform complex tensor processing operations and consume training energy, $E_{\mathrm{train}}$. Once the neural network architecture weights are learned using the data in view of minimizing a loss function, the model is considered ready for evaluation and deployment. Subsequently, the quality of the learned model is evaluated on the evaluation dataset, $N_{\mathrm{S,E}}$, in a process that consumes $E_{\mathrm{eval}}$ energy. The data storage, preprocessing, training, and evaluation form the end-to-end training of the \ac{ai}/\ac{ml} model and cumulatively require $E_{\mathrm{D}}$ energy to complete.

\paragraph{Inference}\label{par:d}
Once the model is trained and evaluated, it can be used by applications or within network scheduling and optimization components in inference mode. $N_{\mathrm{I,P}}$ samples of data can be sent to the AI model running in inference mode and model outputs are returned as responses in the form of forecasts for regression problems, or discrete (categorical) labels for classification problems. The energy consumption $E_{\mathrm{inf}}$ of a single inference is relatively small; however, for high volumes of requests, it can become significant.

\begin{table}[t!]
\centering
\caption{Summary of Notations}
\resizebox{0.73\columnwidth}{!}{%
\begin{tabular}{|c|l|}
\hline
\textbf{Symbol} & \textbf{Description} \\ \hline
\multicolumn{2}{|c|}{\textbf{Counts}} \\ \hline
$N_{\mathrm{S}}$ & No. of collected data samples \\ \hline
\textcolor{black}{$N_{\mathrm{L}}$} & \textcolor{black}{No. of independent links} \\ \hline
$N_{\mathrm{dev,cycle}}$ & No. of processing cycles per bit (device) \\ \hline
$N_{\mathrm{gateway,cycle}}$ & No. of processing cycles per bit (gateway) \\ \hline
$N_{\mathrm{S,T}}$ & No. of samples used for training \\ \hline
$N_{\mathrm{S,E}}$ & No. of samples used for evaluation \\ \hline
$N_{\mathrm{S,inf}}$ & No. of samples used for inference \\ \hline
$N_{\mathrm{epochs}}$ & No. of training epochs \\ \hline
$N_{\mathrm{batch}}$ & No. of batches during training \\ \hline
$I_{\mathrm{S}}$ & Sample size \\ \hline
$\gamma$ & No. of inferences over model lifecycle \\ \hline
$B_{\mathrm{T}}$ & No. of transmitting bits \\ \hline
$B_{\mathrm{T,L_{OSI}}}$ & No. of requested bits by the application from the device \\ \hline
$B_{\mathrm{T}_l}$ & No. of bits need to be processed at the $l$-th layer\\ \hline
$M_{l}$ & No. of nodes in the $l$-th layer \\ \hline
$L_{\mathrm{OSI}}$ & No. of OSI layers \\ \hline
$L$ & No. of dense layers \\ \hline
$C_{\mathrm{in}}$ & No. of input channels \\ \hline
$G$ & No. of intervals in the grid \\ \hline
$N_{\mathrm{f}}$ & No. of filters \\ \hline
$N_{\mathrm{head}}$ & No. of heads \\ \hline
$N_{\text{decoder\_blocks}}$ & No. of decoder blocks \\ \hline
$N_{\mathrm{I,P}}$ & No. of samples used for inference \\ \hline
\multicolumn{2}{|c|}{\textbf{Computational Complexity (CC)}} \\ \hline
$M_{\mathrm{C}}$ & CC of decryption \\ \hline
$M^{\mathrm{minmax}}_{\mathrm{DS}}$ & CC of min-max scaling \\ \hline
$M^{\mathrm{norm}}_{\mathrm{DS}}$ & CC of normalization \\ \hline
$M^{\mathrm{GADF}}_{\mathrm{DS}}$ & CC of GADF \\ \hline
$M_{\mathrm{pre}}$ & CC of preprocessing \\ \hline
$M_{\mathrm{MLP}}$ & CC of a an MLP \\ \hline
$M_{\mathrm{MLP,FP}}$ & CC of a forward propagation of MLP \\ \hline
$M_{\mathrm{CONV}}$ & CC of a convolutional layer \\ \hline
$M_{\mathrm{POOL}}$ & CC of a pooling layer \\ \hline
$M_{\mathrm{CNN,FP}}$ & CC of a forward propagation of CNN \\ \hline
$M_{\mathrm{KAN}}$ & CC of a KAN layer \\ \hline
$M_{\mathrm{NLF}}$ & CC of B-spline activation function across all input elements \\ \hline
$M_{\mathrm{KAN,FP}}$ & CC of a forward propagation of KAN \\ \hline
$M_{\mathrm{ATT}}$ & CC of an attention block \\ \hline
$M_{\mathrm{TR}}$ & CC of the decoder in a transformer \\ \hline
$M_{\mathrm{TR,FP}}$ & CC of a forward propagation of a transformer \\ \hline
$M_{\mathrm{model, tot}}$ & CC of training a model \\ \hline
\multicolumn{2}{|c|}{\textbf{Power}} \\ \hline
$P_{\mathrm{T}}$ & Transmitting power \\ \hline
$P_{\mathrm{R}}$ & Receiving power \\ \hline
$P_{\mathrm{dev,cycle}}$ & Power consumption per processing cycle (device)  \\ \hline
$P_{\mathrm{Gateway,cycle}}$ & Power consumption per processing cycle (Gateway)  \\ \hline
$P_{\mathrm{pre}}$ & Power consumption of the processing unit \\ \hline
\multicolumn{2}{|c|}{\textbf{Energy Consumption (EC)}} \\ \hline
$E_{\mathrm{T}}$ & EC of transmitting data from device to the reception side \\ \hline
$E_{\mathrm{R}}$ & EC of receiving the signal at the reception side \\ \hline
$E_{\mathrm{C}}$ & EC of decryption \\ \hline
$E_{\mathrm{R,tot}}$ & EC of receiving the signal and decryption \\ \hline
$E_{\mathrm{DC}}$ & EC of the data collection component \\ \hline
$E_{\mathrm{DC,b}}$ & EC per bit of the data collection component \\ \hline
$E_{\mathrm{pre}}$ & EC of preprocessing \\ \hline
$E_{\mathrm{train}}$ & EC of model training \\ \hline
$E_{\mathrm{eval}}$ & EC of model evaluation \\ \hline
$E_{\mathrm{inf}}$ & EC of inference \\ \hline
$E_{\mathrm{D}}$ & EC of developing the model \\ \hline
$E_{\mathrm{D,b}}$ & EC per bit of developing the model \\ \hline
$E_{\mathrm{inf,p}}$ & EC of the inference process \\ \hline
$E_{\mathrm{inf,p,b}}$ & EC per bit of the inference process \\ \hline
$eCAL_{\mathrm{abs}}$ & EC over the lifetime of an AI model in the system \\ \hline
$eCAL$ & EC per bit over the lifetime of an AI model in the system \\ \hline
\multicolumn{2}{|c|}{\textbf{Time}} \\ \hline
$T_{\mathrm{T}}$ & Transmitting Time \\ \hline
$T_{\mathrm{pre}}$ & Executing time for preprocessing \\ \hline
\multicolumn{2}{|c|}{\textbf{Parameters}} \\ \hline
$f$ & Bit precision \\ \hline
$\gamma_{l}$ & Scaling factor between DP and CP overheads \\ \hline
$\gamma_{\mathrm{v}}$ & Scaling factor for virtualization \\ \hline
$\beta$ & Split ratio between training and evaluating data \\ \hline
$OH_{\mathrm{dp},l}$ & DP overhead of the $l$-th OSI layer \\ \hline
$OH_{\mathrm{cp},l}$ & CP overhead of the $l$-th OSI layer\\ \hline
$RR_{\mathrm{dp},l}$ & DP retransmission rate of the $l$-th OSI layer \\ \hline
$RR_{\mathrm{cp},l}$ & CP retransmission rate of the $l$-th OSI layer\\ \hline
$R_{\mathrm{T}}$ & Transmitting rate \\ \hline
$R_{\mathrm{R}}$ & Receiving rate \\ \hline
$B$ & Batch size \\ \hline
$PU_{\mathrm{performance}}$ & Theoretical peak performance of a processing unit \\ \hline
$M_{\mathrm{PU}}$ & Computational power of a processor \\ \hline
$I_{\mathrm{r}}$ & Input tensor size (height) \\ \hline
$I_{\mathrm{c}}$ & Input tensor size (width) \\ \hline
$K_{\mathrm{r}}$ & Kernel (filter) height \\ \hline
$K_{\mathrm{c}}$ & Kernel (filter) width \\ \hline
$P_{\mathrm{r}}$ & Padding size (height) \\ \hline
$P_{\mathrm{c}}$ & Padding size (width) \\ \hline
$S_{\mathrm{r}}$ & Stride value (height) \\ \hline
$S_{\mathrm{c}}$ & Stride value (width) \\ \hline
$K$ & Spline order \\ \hline
$C$ & Context length \\ \hline
$N_{\mathrm{embed}}$ & Size of the embedding \\ \hline
FFS & Feed forward size \\ \hline
\end{tabular}%
}
\label{tab:notations}
\vspace{-25pt}
\end{table}

\subsection{Existing Metric Relevance for the Data Manipulating Components of the AI Lifecycle}

\textcolor{black}{Existing energy metrics offer only partial visibility into the energy demands of \ac{ai} models across their full lifecycle, as illustrated in Table~\ref{tab:3gpp}. On the one hand, telecommunication and data center metrics (columns~2–6) primarily address the data collection stage. Telecommunication metrics focus on physical-layer energy consumption in the access network, while data center metrics capture aggregate energy use from all IT infrastructure. However, these metrics lack formal attribution to specific network segments or \ac{osi} layers, limiting their granularity and diagnostic utility. On the other hand, the next four columns (columns~7–10) show that existing deep learning metrics provide only coarse coverage of energy consumption during data preprocessing, training, evaluation, and inference. These metrics are typically limited to classification tasks and often rely on weight counts as proxies for energy use. Importantly, only two of them explicitly consider inference, underscoring a significant gap in coverage. Therefore, there is a need for the development and adoption of the proposed eCAL metric (column~11) as a unified, lifecycle-aware measure that enables consistent and component-resolved evaluation of energy efficiency in AI systems.}
 
\subsection{Proposed Metric} 
Based on the AI model lifecycle illustrated in Figure \ref{fig:system_model} and discussed in this section, we propose eCAL as \textit{the ratio of total energy consumed by data manipulation components to the total manipulated application-level bits}: 

\begin{small}
    \begin{align}
    eCAL=\frac{\text{Total energy of data manipulation components } [J]}{\text{Total manipulated application level data } [b]}.
\label{eq:e_inf_tot_b}
\end{align}
\end{small}

\textcolor{black}{eCAL introduces a fundamentally new way of looking and assessing the cost of using AI - it essentially enables to quantify the energy cost of each bit that comes out from the inference of an AI model over its entire lifetime while accounting for the underlying overheads. The purpose of eCAL is to measure the overall energy consumption of any AI model developed using data and a neural network architecture, independent of the use case and accuracy/performance expected by that use case, similarly as for instance, \ac{ee} from Table \ref{tab:3gpp} is applicable to various communication modules independent on their spectral efficiency. As illustrated in Table \ref{tab:3gpp}, eCAL is the only metric able to capture the energy costs in communication networks following the standard OSI model by  modeling the overheads introduced by various layers as well as two of the layers of the standard CCRA in cloud computing. All the other metrics have a narrower scope and do not center on and include the actual AI training and inference. In future when a metric such as eCAL may be standardized, it could be used as an energy badge on any AI model, in a similar fashion as there are CO$_{2}$ footprint badges used when booking flights or \ac{ee} classes used on household appliances. }

\subsection{Methodology} 
In the upcoming sections, we derive the proposed metric for  example case studies that comply with the OSI \textcolor{black}{standard architecture} and MLOps framework that is part of the platform layer in the \textcolor{black}{CCRA standard architecture}. We follow the data flow, highlighted with blue in Fig.~\ref{fig:system_model}, across the above-mentioned components and derive the computational complexity and subsequently the energy consumption, highlighted with purple, in each component. In Section \ref{sec:E_T}, we derive $E_{\mathrm{T}}$ for the data collection component. With this formalism in place, Section \ref{sec:e2e} is concerned with deriving eCAL under specific configuration assumptions, without limiting the scope of the proposed metric. For the sake of clarity and convenience, we summarize the notations used throughout this paper in Table~\ref{tab:notations}.

\section{Energy cost of Data Collection}
\label{sec:E_T}

In this section, we derive the energy cost of \textit{data collection} in an \ac{ai}-powered wireless system. To calculate this cost, we first determine the number of bits that need to be transmitted from a device to the server via wireless technology (uplink). Then, we calculate the energy consumption of transmitting and receiving the data, based on the given transmission and reception power and the corresponding data rate. Finally, we derive the total energy consumption for collecting data from a device. 
The assumption is that all data is collected via a single link; however, a summation can be added to generalize data collection to multiple links with different parameters. In this regard, the modular and extensible open source calculator can be extended to include specific communication segments, topologies, and technologies employing various resource allocation and optimization techniques.
\subsection{\ac{osi} Overhead Estimation Method in eCAL}

 In \ac{osi} architecture, each layer introduces its own overhead, including various header fields in the \ac{dp} and additional control and set-up messaging from the \ac{cp}. Additionally, in some of the layers, such as link (e.g., MAC) and transport (e.g., TCP) layers, errors or packet loss are internally treated, potentially leading to retransmissions. While each protocol and every layer enables a wide selection of settings, we devise the following generalized theoretical formulation for computing the size of the transmitted data at the physical layer ($B_{\mathrm{\mathrm{T}}}$):
\begin{align}
     B_{\mathrm{T}}&~[b] =\lceil B_{\mathrm{T,L_{\mathrm{OSI}}}} \prod_{l=1}^{L_{\mathrm{OSI}}} \underbrace{\left(RR_{\mathrm{DP},l}  (1 + OH_{\mathrm{DP},l})\right.}_{\text{contribution from DP} }\nonumber\\
     &+ \underbrace{\left.RR_{\mathrm{CP},l} \gamma_l   OH_{\mathrm{CP},l}\right)}_{\text{contribution from CP}} \rceil,~\mathrm{where}~L_{\mathrm{OSI}}=7,
     \label{eq:bits}
\end{align}

\noindent where $B_{\mathrm{T,L_{OSI}}}$ represents the application layer data (e.g. telemetry) sent from the device, measured in bits. It is calculated as the product of the bit precision ($f$), the number of \textit{samples} ($N_{\mathrm{S}}$) and the \textit{sample size} ($I_{\mathrm{S}}$) in floating-point format, i.e., $B_{\mathrm{T,L_{\mathrm{OSI}}}}=f N_{\mathrm{S}}I_{\mathrm{S}}$. Typically, $f$ equals $32$ bits for single precision (float) or 64 bits for double precision format. To be able to send/receive data, in each of the seven \ac{osi} layers, represented by the product in Eq.~(\ref{eq:bits}), a number of data and control plane operations take place as expressed in the respective terms. The application data, $B_{\mathrm{T,L_{OSI}}}$, is processed and is added some overhead $OH_{\mathrm{DP},l}$, in some layers it is encrypted (i.e. application, presentation and/or transport layers) and in some layers it may also require retransmissions $RR_{\mathrm{DP},l}$. Moreover, some layers, such as the network layer, incur additional overhead required from signaling to maintain routing tables. This is reflected in the second term of Eq.~(\ref{eq:bits}) through the respective \ac{cp} overheads $OH_{\mathrm{CP},l}$, retransmissions $RR_{\mathrm{CP},l}$ and the $\gamma_{l}$ scaling factor that adjusts for the relative contribution of the \ac{cp} overhead compared to the \ac{dp} counterpart at the $l$-th layer. The scaling factor can also be used to compensate for other specific overheads unforeseen by the considered configuration in this paper, \textcolor{black}{such as the fiber-based wired communication backhaul.} 

\paragraph*{\textcolor{black}{Numerical Validation}} The eCAL calculator implements the computation from Eq.~(\ref{eq:bits}) and enables the configuration of the application data size, retransmission rates, overheads, and scaling factor. Furthermore, it enables plugging custom protocol implementations for each of the layers, based also on the existing literature~\cite{TCP_energy,Application_energy,IPV6_energy}, so that the community can also instantiate eCAL for a specific well-defined set of protocols in addition to the approach adopted in this paper.

\begin{figure}[t!]
    \centering
    \includegraphics[width=0.48\textwidth]{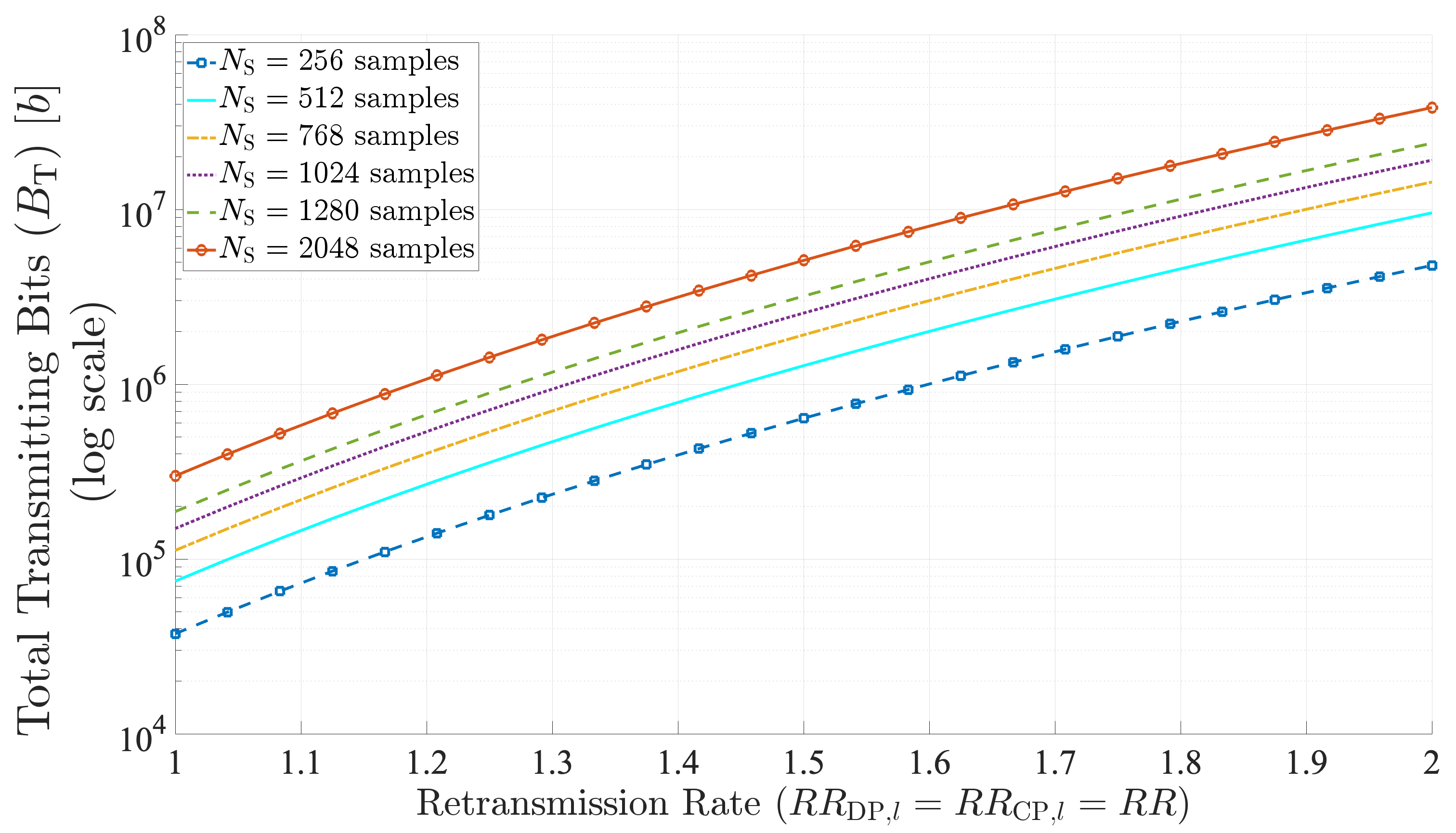}
    \caption{Total transmitting bits ($B_{\mathrm{T}}$) versus retransmission rate with $\gamma_{l}=0.5$, $OH_{\mathrm{DP},l}=10\%$ and $OH_{\mathrm{CP},l}=5\%$.}
    \label{fig:bitvsrr}
    \vspace{-10pt}
\end{figure}

To illustrate the behavior of Eq.~(\ref{eq:bits}), we first consider a scenario where a single device transmits $N_{\mathrm{S}}$ of double-precision samples to the application accompanied by \ac{dp} and \ac{cp} overheads while altering the retransmission rate. In particular, we consider that each layer in the \ac{osi} model introduces a fixed overhead percentage for the \ac{dp} and \ac{cp}, with $10$\% and $5$\%, respectively. Furthermore, the relative contribution of \ac{cp} overhead is scaled by a factor of $0.5$, i.e., $\gamma_{l}=0.5$. The retransmission rates for both \ac{dp} and \ac{cp} are assumed to be equal and vary from $1$ to $2$ in increments of $0.04$. This is shown in Fig.~\ref{fig:bitvsrr}.  The results show the retransmission rate has a significant impact on total transmitting bits. For example, when $N_{\mathrm{S}}=256$ samples, there is a growth of $2.3$, $17$, and $128$ times with retransmission rates equal to $1.125$, $1.5$, and $2$, respectively, compared to scenarios with no retransmission. 
\subsection{\textcolor{black}{Application Layer Centered Energy Estimation for the Data Collection Component of eCAL}}
To estimate the energy consumption of the data collection component ($E_{\mathrm{DC}}$), which is the sum of the energy of transmitting ($E_{\mathrm{T}}$) and receiving the data ($E_{\mathrm{R}}$) introduced in the \ac{phy}, as well as the energy consumption of computation in the \ac{dp} and \ac{cp} across the various layers of \ac{osi} ($E_{\mathrm{C}}$). \textcolor{black}{We consider a network comprising $N_{\mathrm{L}}$ independent and direct links between nodes, which covers all the network topologies, including multi-access and multi-hop links. We propose the following analytical expression for the $p$-th links, which includes:}
\textcolor{black}{
\begin{equation}
\small
\begin{aligned}
E_{\mathrm{DC},p}&= \\
&\underbrace{\frac{P_{\mathrm{T},p}[W]}{R_{\mathrm{T},p}[b/s]}  B_{\mathrm{T},p}[b]}_{\text{PHY layer transmitting energy ($E_{\mathrm{T},p}$)}}+\underbrace{\frac{P_{\mathrm{R},p}[W]}{R_{\mathrm{R},p}[b/s]}  B_{\mathrm{T},p}[b]}_{\text{PHY layer receiving energy ($E_{\mathrm{R},p}$)}} \\
  +\sum^{L_{\mathrm{OSI}}}_{l=2}&(\underbrace{B_{\mathrm{T},l,p}~[b]   N_{\mathrm{dev, cycle},l,p}~[c/b]  P_{\mathrm{dev, cycle},p}~[J/c]}_{\text{from MAC to APP layer computational energy at the  device}}\\
 +&\underbrace{B_{\mathrm{T},l,p}~[b]   N_{\mathrm{gateway, cycle},l,p}~[c/b]  P_{\mathrm{gateway,cycle},p}~[J/c]}_{\text{from MAC to APP layer computational energy at the gateway}}),
\end{aligned}
\label{eq:E_DC_final} 
\end{equation}
}

\noindent where $P_{\mathrm{T},p}$ and $R_{\mathrm{T},p}$ represent the transmitting power and the data rate, respectively, while $P_{\mathrm{R},p}$ and $R_{\mathrm{R},p}$ are the receiving power and data rate. Furthermore, $B_{\mathrm{T},l}$ denotes the total number of bits that need to be processed at the $l$-th layer in the \ac{osi} model. Additionally, we define the computational parameters at the device as $N_{\mathrm{dev, cycle},l,p}$ representing the number of processing cycles per bit at the $l$-th layer, and $P_{\mathrm{dev, cycle}}$, denoting the power consumption per cycle. Similarly, at the gateway, these parameters are defined as $N_{\mathrm{gateway, cycle},l}$ and $P_{\mathrm{gateway, cycle}}$. \textcolor{black}{As a result, we can calculate the total energy cost of the data collection as the sum of all $N_{\mathrm{L}}$ links, which is expressed as:
\begin{equation}
E_{\mathrm{DC}}=\sum^{N_{\mathrm{L}}}_{p=1}E_{\mathrm{DC},p}.
\label{eq:E_DC_sum}
\end{equation}
}
\begin{figure}[t]
    \centering
    \includegraphics[width=0.48\textwidth]{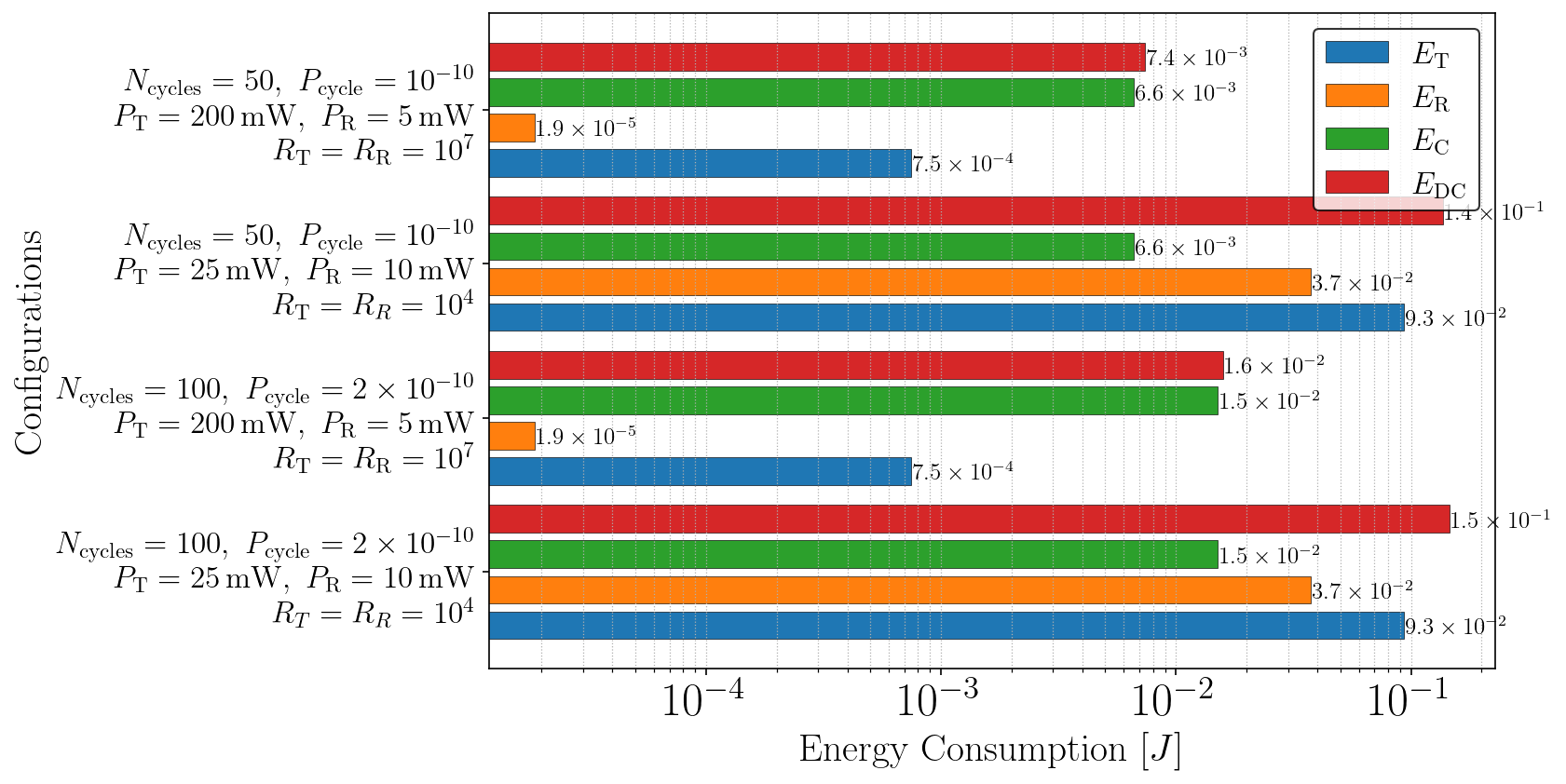}
    \caption{Energy consumption of data collection ($E_{\mathrm{DC}}$) (log scale) with its components and configurations, including transmission energy ($E_{\mathrm{T}}$), receiving energy ($E_{\mathrm{R}}$), and computational energy ($E_{\mathrm{C}}$) with $N_{\mathrm{S}}=256$, $RR_{\mathrm{DP},l}=RR_{\mathrm{CP},l}=1$, $OH_{\mathrm{DP},l}=10\%$, $OH_{\mathrm{CP},l}=5\%$, and $\gamma_{l}=0.5$.}
    \label{fig:E_DC_tot}
    \vspace{-10pt}
\end{figure}
\paragraph*{\textcolor{black}{Numerical  Validation}} Computing exact values for $E_{\mathrm{DC}}$ for a broad range of $B_{\mathrm{T}_l}$, $N_{\mathrm{\_, cycle},l}$, and  $P_{\mathrm{\_, cycle}}$ is challenging due to the overall complexity of computing systems. For instance, $N_{\mathrm{\_, cycle,}l}$ varies across microprocessors when processing the same amount of data,  and it does not have a linear dependency with data size~\cite{AES_paper}. Therefore, the values for $B_{\mathrm{T}_l}$ and $N_{\mathrm{\_, cycle}}$ cannot be chosen arbitrarily. Next, $P_{\mathrm{\_, cycle}}$ also depends on the microprocessor. However, in Fig.~\ref{fig:E_DC_tot}, we illustrate the energy consumption of $E_{\mathrm{DC}}$ with a selected set of values to provide an intuition on the contribution from each term. More specifically, we analyze a scenario where $N_{\mathrm{S}}=256$, and we consider two sets of parameters to compute transmitting and receiving energy, i.e. $E_{\mathrm{T}}$ and $E_{\mathrm{R}}$, alongside two different settings for calculating the computational energy consumption ($E_{\mathrm{C}}$). The results show that while transmission energy ($E_{\mathrm{T}}$) remains the dominant contributor in most cases, the computational energy consumption ($E_{\mathrm{C}}$) is not negligible and, in some configurations, constitutes a significant portion of the total energy consumption of data collection. This is particularly the case when the hardware is less optimized for the tasks~\cite{AES_paper}, leading to higher cycles per bit ($N_{\mathrm{\_, cycle},l}$) and power per cycle ($P_{\mathrm{\_, cycle}}$). For example, when $N_{\mathrm{\_,cycle}}=100$ and $P_{\mathrm{\_, cycle}}=2\times 10^{-10}$, $E_{\mathrm{C}}$ is $1.51\times 10^{-2}~[J]$, which represents approximately $10\%$ and $95\%$ of the total energy cost in two different \ac{phy} configurations. On the other hand, when $N_{\mathrm{\_,cycle},l}=50$ and $P_{\mathrm{\_, cycle}}=10^{-10}$, $E_{\mathrm{C}}$ is $6.61\times 10^{-3}~[J]$, contributing to approximately $5\%$ and $90\%$ of the total energy consumption in data collection component.
\begin{figure}[t]
    \centering
    \includegraphics[width=0.48\textwidth]{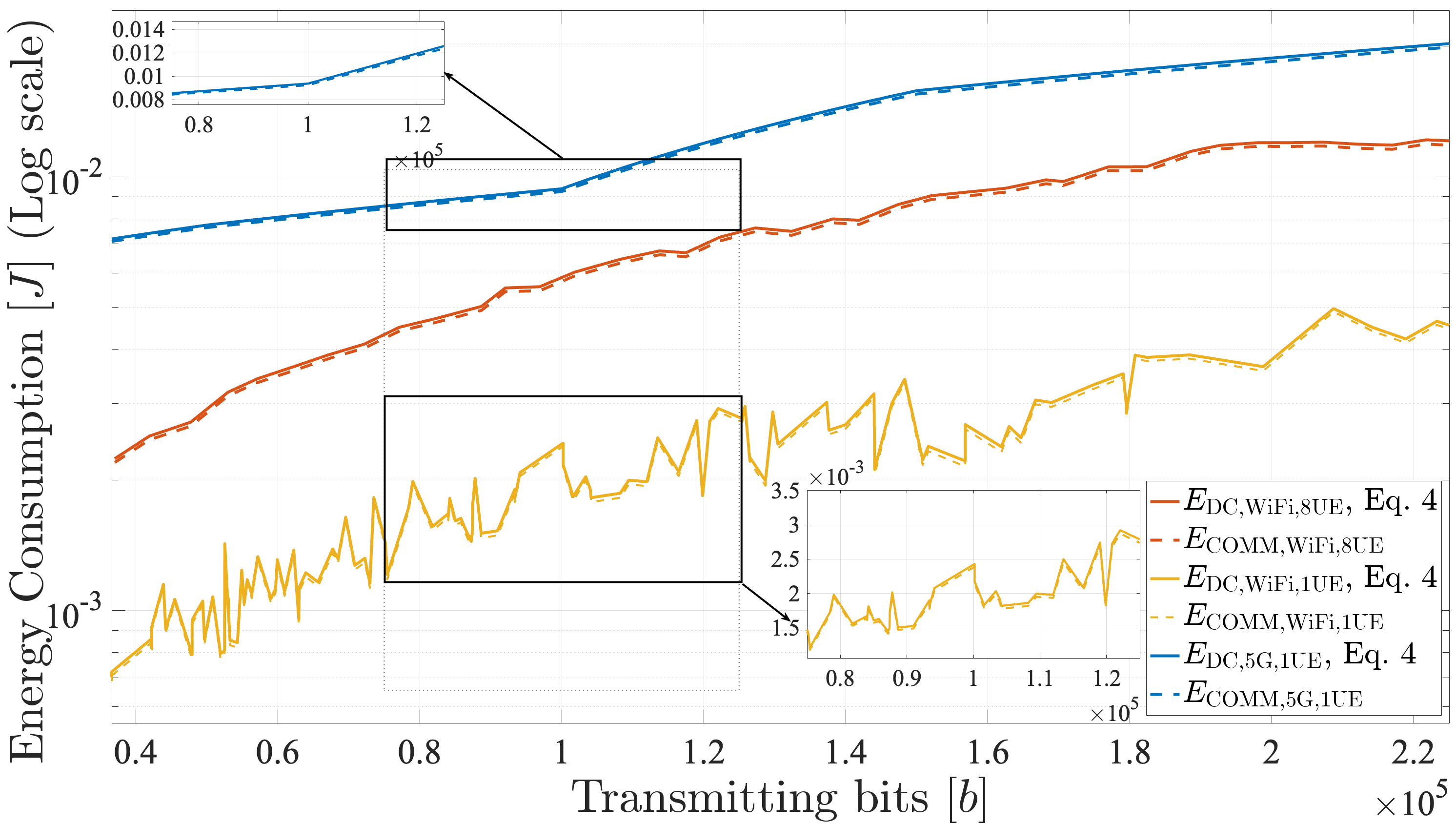}
    \caption{Proposed $E_{\mathrm{DC}}$ comparison with real-world measurement data.}
    \label{fig:e_dc}
    \vspace{-10pt}
\end{figure}

\begin{table}[!h]
\centering
\caption{Comparison of \ac{ee} when transmitting $0.1$ Mbits}
\label{tab:ee_dc_comparison}
\footnotesize 
\begin{tabular}{lcc}
\toprule
\textbf{Tech} & \textbf{EE w/ $E_{\mathrm{DC}}$ (Mbit/J)} & \textbf{EE (Mbit/J)} \\
\midrule
Wi-Fi (8 UEs) & 17.07 & 17.42 \\
Wi-Fi (1 UE)  & 41.31 & 41.32 \\
5G            &  1.63 &  1.65 \\
\bottomrule
\end{tabular}
\end{table}

\paragraph*{\textcolor{black}{Empirical Validation}} \textcolor{black}{To further validate our approach in a real-world setting, empirical measurements from Wi-Fi and 5G networks are considered. Specifically, in~\cite{wifi_DC}, the authors conducted rigorous measurements of various parameters, such as \ac{ofdma} throughput, latency, and power consumption achieved by Wi-Fi 6 technology, i.e., IEEE 802.11ax, using commercial devices. For 5G measurements, we consider the experiment conducted in~\cite{5g_dc}, where a mobile \ac{ue} is connected to a commercial 5G modem.}

\textcolor{black}{The results are shown in Fig.~\ref{fig:e_dc}. The proposed $E_{\mathrm{DC}}$ consistently serves as a tight upper bound across both technologies, and hence corresponds to a lower bound on achievable \ac{ee} (bits per Joule). For example, under the same bit window size of $0.75$ to $1.2\times 10^5$ [$b$] the measured energy consumption ($E_{\mathrm{COMM,\_}}$) of 5G (highlighted in blue), Wi-Fi 6 in a multi-access scenario ($8$ \acp{ue}, highlighted in red), and single \ac{ue} (highlighted in orange) are $8.4$ to $12.4$ [$mJ$], $4.1$ to $7.2$ [$mJ$], and $1.4$ to $2.75$ [$mJ$], respectively. The proposed approach to the corresponding cases are $9.38$ to $15.8$ [$mJ$], $6.03$ to $8.63$ [$mJ$], $1.82$ to $3.41$ [$mJ$]. Moreover, the comparison of the \ac{ee} when transmitting $0.1$ Mbits of data can be seen in Table~\ref{tab:ee_dc_comparison}. As discussed, the proposed approach serves as a tight lower bound for \ac{ee}. In other words, if we put the energy consumption of the computational part into consideration, this data-manipulating component will be less energy efficient than only considering physical layer measurement.} 

\textcolor{black}{The methodology used for obtaining these results is as follows. We aggregate and process three datasets that are collected in real-world measurements, corresponding to uplink with $8$ \acp{ue} (1 mobile \ac{ue} and 7 \acp{pc} with Wi-Fi), uplink with a single mobile \ac{ue} (Wi-Fi), and single \ac{ue} in a commercial 5G network. For the Wi-Fi scenarios, we extract device-side transmission energy and device specifications from measurements to parameterize the \ac{ue} terms in Eq.~(\ref{eq:E_DC_final}); for the \ac{ap} (ASUS RT-AX58U), we use hardware specifications to obtain the receiving and computing terms in Eq.~(\ref{eq:E_DC_final}). To model the \ac{osi}‐stack overhead, we use published cryptography and networking guidance. On mobile \ac{ue} (Galaxy S10, ARMv8 with cryptography extensions)~\cite{ARMv8CryptoExt}, we assign $0.3$ [$c/b$] from transport to application layer (L4 to L7) and $0.1$ [$c/b$] at the \ac{mac} and Network layer (L2 and L3). These numbers are consistent with \ac{aes}-based encryption throughput on ARM-based architectures with hardware acceleration reported in OpenSSL microbenchmarks~\cite{OpenSSLSpeed}
. On PCs (Intel AX210 NICs) and \ac{ap}, link-layer security and \ac{mac} functions are \ac{nic}-offloaded. 
Therefore, we assume $0.02$ [$c/b$] (application), $0.01$ [$c/b$] (TCP/IP protocol), and $0.005$ [$c/b$] (MAC), aligned with Intel AES documentation and NIC offload literature~\cite{IntelAESNI,IntelVAES}. Guided by~\cite{Horowitz2014ISSCC}, we set $0.3$ [$nJ/c$] and $0.1$ [$nJ/c$] on ARM and Intel-based systems, respectively. For 5G measurement, we can obtain the receiving power and the receiving data size. It is mentioned in~\cite{AIOT_00} that the pathloss follows the 3GPP UMa model, so we can then obtain the transmitting power. To account for protocol stack overhead, we set the overhead for L2 to L6 as $0.2$ and L7 as $0.1$. Since the setup of \ac{ue} and \ac{ap} is similar to the Wi-Fi scenario (same UE and Intel-based processor), the computing parameters are assumed to be the same as in the Wi-Fi scenario.  
These parameters help to anchor the derived $E_{\mathrm{DC}}$ to realistic hardware capabilities at practical operating points.
}

\section{Energy Cost of Data Preprocessing}
\label{sec:preproc}
At the server side, data is first stored on a hard drive, subsequently preprocessed, and finally used for the \ac{ml} model development or inference. This section focuses on preprocessing, while inference is addressed in Section~\ref{sec:E_inf}. 

Storage is generally energy-efficient, with relatively negligible energy costs. However, it can become significant when handling large volumes of incoming data or when utilizing virtualized hosted storage with encryption in an edge-cloud environment~\cite{cloud_storage}. In the following case study, we consider the dataset with $N_{\mathrm{S}}$ samples collected via the wireless network, as depicted in Fig.~\ref{fig:system_model} and discussed in Section \ref{sec:E_T}. We also consider data preprocessing consisting of \textit{data cleaning}, and \textit{data transformation} steps, while leaving other more complex approaches, such as feature engineering, for further extensions of the study and simulation tool. 

In the \textit{data cleaning} step, the task of preprocessing is to identify and remove invalid samples from a list or array, which incurs zero \ac{FLOPS}\footnote{FLOPs are often used to enable theoretical estimation of computational complexity. Alternative approaches would involve direct energy measurements through interfaces such as PAPI or tools such as Scaphandre.}, as no floating-point additions or multiplications are performed, only comparisons and memory operations. During the \textit{data transformation} step, preprocessing involves determining the complexity of data transformation ($M_{\mathrm{DS}}$). We provide three examples, i.e., min-max scaling, normalization, and \ac{gadf} process, which represent typical approaches for data transformation.

\paragraph{Min-max scaling}
Min-max scaling is a data preprocessing technique that transforms numerical data to fit within the [0,1] interval by scaling each value relative to the minimum and maximum values in the dataset. For each feature, the transformation subtracts the minimum value and divides by the range, ensuring that the lowest value becomes 0 and the highest value becomes 1, with all other values scaled proportionally between them. While finding the minimum and maximum values requires computational resources for comparisons, these operations do not contribute to the \ac{FLOPS} count. The scaling itself requires calculating the range (one FLOP), followed by subtracting the minimum and dividing by the range for each sample ($2  N_{\mathrm{S}}  I_{\mathrm{S}}$ \ac{FLOPS}). Therefore, the total number of \ac{FLOPS} ($M^{\mathrm{minmax}}_{\mathrm{DS}}$) for min-max scaling can be expressed as:
\begin{equation}
\begin{aligned}
M^{\mathrm{minmax}}_{\mathrm{DS}} = 2  N_{\mathrm{S}}  I_{\mathrm{S}} +1.
\end{aligned}
\end{equation}
\paragraph{Normalization}

Standard score, also known as z-score normalization, represents a more computationally intensive preprocessing technique that involves three distinct computational steps. First, calculating the population mean requires $N_{\mathrm{S}}  I_{\mathrm{S}}$ \ac{FLOPS}, as we sum all values and divide by the total number of samples. Second, computing the population standard deviation demands $3  N_{\mathrm{S}}  I_{\mathrm{S}} + 1$ \ac{FLOPS}, which encompasses calculating squared differences from the mean, their sum, and the final square root operation. Finally, the standardization transformation necessitates $2  N_{\mathrm{S}}  I_{\mathrm{S}}$ \ac{FLOPS}, as each data point must be centered by subtracting the mean and divided by the standard deviation for scaling. The total computational cost for normalization, expressed in \ac{FLOPS} ($M^{\mathrm{norm}}_{\mathrm{DS}}$), is therefore:

\begin{equation}
\begin{aligned}
M^{\mathrm{norm}}_{\mathrm{DS}} = 6  N_{\mathrm{S}}  I_{\mathrm{S}} + 1.\label{eq:flops_pre_normalization}
\end{aligned}
\end{equation}

\paragraph{Gramian Angular Difference Field}

To showcase an even more computationally expensive preprocessing operation, we implemented a calculator to compute the energy cost of the \ac{gadf} transformation~\cite{ref43_pub}. This transformation converts time series data into an image representation that captures temporal correlations between each pair of values in the time series. The \ac{gadf} computation consists of several sequential steps, each contributing to the overall computational complexity and, consequently, the energy consumption.
The first step requires scaling the time series data using min-max scaling, which ensures that values fall within the [-1, 1] interval. Following the scaling, the data undergoes conversion to polar coordinates, where the radius is calculated using the time stamp values, and the angular coordinates are computed through an inverse cosine operation. The final step involves constructing the \ac{gadf} matrix through pairwise angle differences, resulting in an $I_{\mathrm{S}} \times I_{\mathrm{S}}$ matrix where $I_{\mathrm{S}}$ represents the length of the input time series. 

The computational complexity of \ac{gadf} can be broken down into its constituent operations. The initial min-max scaling requires $2  N_{\mathrm{S}}  I_{\mathrm{S}} + 1$ \ac{FLOPS} as previously discussed. However, in this case we have to multiply the number of samples with the length of the input time series ($I_{\mathrm{S}}$) as we are dealing with a series of values in a single sample increasing the \ac{FLOPS}. The polar coordinate transformation and the construction of the \ac{gadf} matrix requires $(5  I_{\mathrm{S}} + I_{\mathrm{S}}^2)   N_{\mathrm{S}}$ \ac{FLOPS} . Therefore, the total number of \ac{FLOPS} ($M^{\mathrm{GADF}}_{\mathrm{DS}}$) for the \ac{gadf} transformation can be expressed as:
\begin{equation}
\begin{aligned}
M^{\mathrm{GADF}}_{\mathrm{DS}} = \left(2  N_{\mathrm{S}}  I_{\mathrm{S}} + 1\right) + (5  I_{\mathrm{S}} + I_{\mathrm{S}}^2)  N_{\mathrm{S}}.
\label{eq:gadf_equation}
\end{aligned}
\end{equation}
This quadratic computational complexity makes \ac{gadf} significantly more energy-intensive compared to simpler preprocessing operations like standardization or min-max scaling, particularly for longer time series. As a result, the total number of \ac{FLOPS} for preprocessing can be expressed as:
\begin{equation}
M_{\text{pre}} = 
\begin{cases}
    2  N_{\text{S}}  I_{\mathrm{S}} + 1,  \text{min-max scaling}, \\
    6  N_{\text{S}}  I_{\mathrm{S}} + 1,  \text{normalization}, \\
    2  N_{\mathrm{S}}  I_{\mathrm{S}} + 1 +(5  I_{\mathrm{S}} + I_{\mathrm{S}}^2)  N_{\mathrm{S}}, \text{GADF}. \label{eq:pre_all_complexity}
\end{cases}
\end{equation}
 
With the computing complexity in \ac{FLOPS} formalized, we can obtain total energy consumption ($E_{\mathrm{pre}}$) of the preprocessing ($E_{\mathrm{pre}}$) by utilizing the power consumption of the processing unit ($P_{\mathrm{pre}}$) and the executing time ($T_{\mathrm{pre}}$): 
\begin{equation}
    E_{\mathrm{pre}}[J]= P_{\mathrm{pre}}[W]  T_{\mathrm{pre}}[s],\label{eq:e_pre}
\end{equation}
where $T_{\mathrm{pre}}$ can be calculated as the ratio between FLOPs needed for preprocessing and the computational power of a processor ($M_{\mathrm{PU}}$): 
\begin{equation}
T_{\mathrm{pre}}[s] = \frac{M_{\mathrm{pre}}[FLOPs]}{M_{\mathrm{PU}}[FLOPs/s]}.\label{eq:T_pre}
\end{equation}


\begin{figure}[t]
    \centering
    \includegraphics[width=0.48\textwidth]{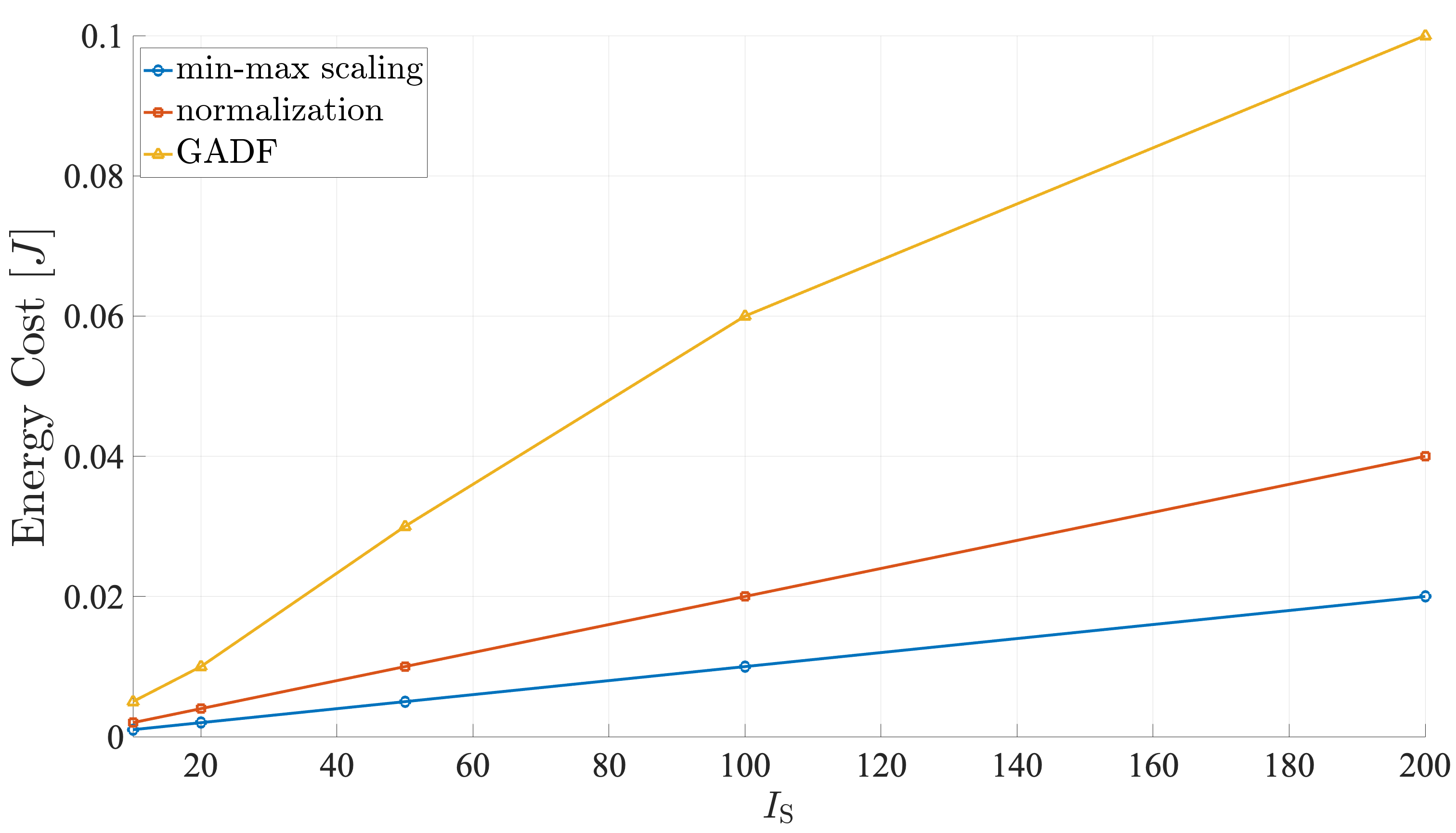}
    \caption{Energy consumption ($E_{\mathrm{pre}}$) of different sample size ($I_{\mathrm{S}}$) across different preprocessing techniques.}
    \label{fig:energy_different_data_size_pre}
    \vspace{-10pt}
\end{figure}
\begin{figure}[t]
    \centering
    \includegraphics[width=0.48\textwidth]{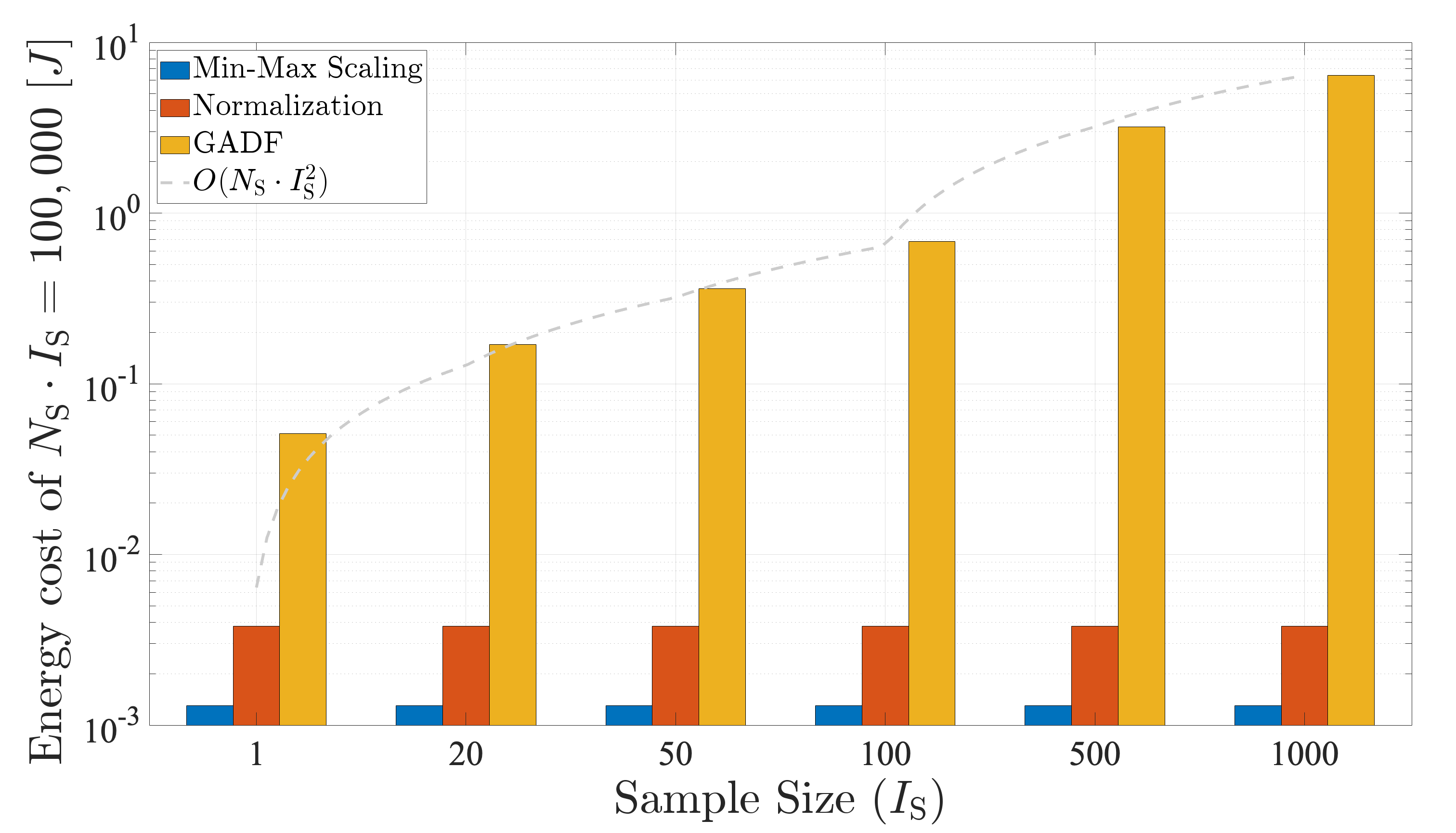}
    \caption{Energy consumption ($E_{\mathrm{pre}}$) of fixed data point ($N_{\mathrm{S}}  I_{\mathrm{S}}$) across different preprocessing techniques and sample size ($I_{\mathrm{S}}$).}
    \label{fig:energy_different_I_S}
    \vspace{-10pt}
\end{figure}

\paragraph*{\textcolor{black}{Numerical Validation}} To illustrate the energy cost in various $N_{\mathrm{S}}$ and $I_{\mathrm{S}}$ configurations, consider the preprocessing energy cost $E_{\mathrm{pre}}$ on a CPU with $M_{\mathrm{PU}}=10~\text{GFLOPs/s}$ and power consumption of $P_{\mathrm{pre}}=140~\text{W}$, shown in Fig.~\ref{fig:energy_different_data_size_pre} and Fig.~\ref{fig:energy_different_I_S}. More specifically, Fig.~\ref{fig:energy_different_data_size_pre} demonstrates how the energy consumption of preprocessing methods for a fixed number of samples ($N_{\mathrm{S}}=256$) increases with the sample size ($I_{\mathrm{S}}$).  This trend is consistent across min-max scaling, normalization, and \ac{gadf}, with \ac{gadf} exhibiting the most pronounced energy growth due to its quadratic dependency on $I_{\mathrm{S}}$, whereas, the energy cost of the other two techniques grows linearly. In addition, Fig.~\ref{fig:energy_different_I_S} focuses on the energy consumption of the preprocessing techniques while maintaining a constant total amount of data ($N_{\mathrm{S}}   I_{\mathrm{S}} = 100,000$). This constraint ensures that as the sample size ($I_{\mathrm{S}}$) increases, the number of data samples ($N_{\mathrm{S}}$) decreases proportionally. Consequently, the energy consumption for min-max scaling and normalization remains constant across different configurations, as their computational complexity scales linearly with $N_{\mathrm{S}}   I_{\mathrm{S}}$. In contrast, \ac{gadf} exhibits a quadratic (with a linear component) growth in energy consumption due to the $I^{2}_{\mathrm{S}}$ term in its computational complexity, shown in Eq.~(\ref{eq:pre_all_complexity}), which is also confirmed by the $O\left(N_{\mathrm{S}}   I^{2}_{\mathrm{S}}\right)$ shown in the gray dotted line. These results highlight the trade-offs in selecting a preprocessing method based on both the \ac{ee} requirements and the specific input data characteristics required by the \ac{ai}/\ac{ml} technique.
\section{Energy Cost of Training}
\label{sec:e_train}
To train a model, training data and machine learning techniques are needed. In this study, we consider neural network architectures with centralized offline training, however, extensions of this work could consider online, distributed, federated or transfer approaches as well. As discussed in Section \ref{sec:preproc}, the training data typically goes through a preprocessing phase and is then split into training and evaluation sets, with a $\beta$ split ratio as:
\begin{equation}
    N_{\mathrm{S}}=\underbrace{\beta   N_{\mathrm{S}}}_{N_{\mathrm{S,T}}}+\underbrace{(1-\beta)  N_{\mathrm{S}}}_{N_{\mathrm{S,E}}},
\end{equation}

\noindent where $N_{\mathrm{S,T}}$ and $N_{\mathrm{S,E}}$ represent the number of samples for training and evaluation, respectively. \textcolor{black}{For studying the energy cost of training, in this paper we focus on four representative neural network architectures\footnote{Note that the implementation of the calculator supports many other architectures using pytorch.},  namely, \acp{mlp}~\cite{GARCIAMARTIN201975}, \acp{cnn}~\cite{cnn_ref}, \acp{kan}~\cite{KAN_intro}, and transformers~\cite{vaswani2017attention}. MLPs established the foundations of deep learning as universal function approximators, enabling the first wave of practical neural network models. CNNs marked a breakthrough by exploiting spatial hierarchies, revolutionizing computer vision and inspiring domain-specific architectures. Transformers represent the most recent paradigm shift, introducing attention mechanisms that now dominate language processing through the GPT models. Finally, KANs are included as a novel, explainable architecture compared to MLPs.}

\begin{figure}[!thb]
    \centering
    \includegraphics[width=0.48\textwidth]{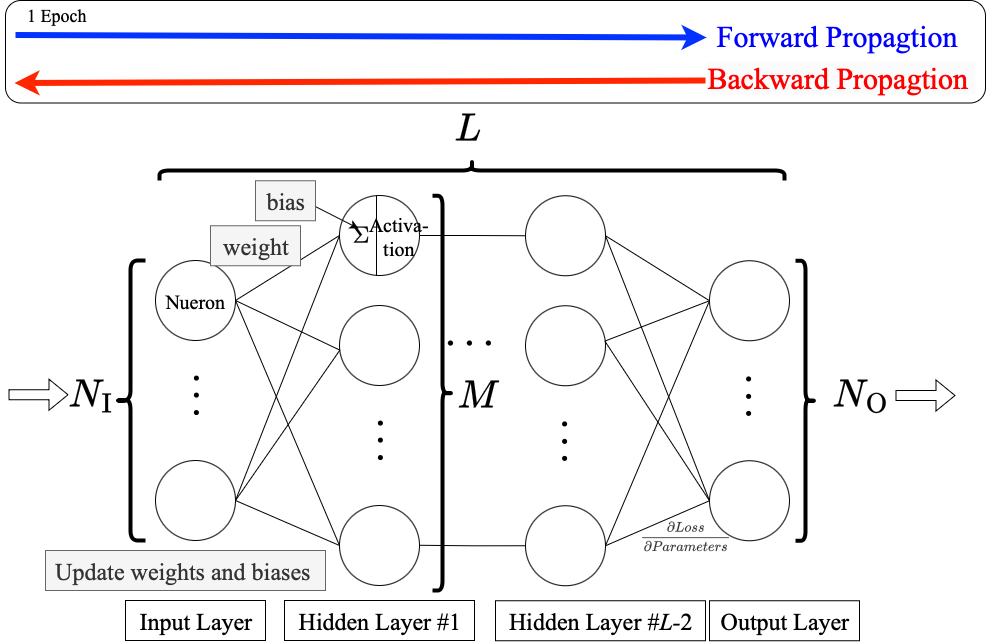}
    \caption{A fully connected MLP architecture.}
    \label{fig:MLP_model}
\end{figure}

Training a neural network involves two fundamental phases: \textit{forward propagation} and \textit{backward propagation}, depicted with blue and red arrows in Fig.~\ref{fig:MLP_model}, respectively, on an example MLP. During the forward propagation, input data passes through the network layers. Each layer performs specific tasks, such as computing a linear combination of weights and biases, followed by applying activation functions in dense (fully connected) layers. Convolutional layers perform convolutional operations, while pooling mechanisms are applied on pooling layers. At the end of this phase, the loss is calculated by comparing the predicted outputs to the actual target values. Subsequently, the model performs backward propagation, where it calculates gradients of the loss with respect to each parameter, updates these gradients in reverse order through the layers, and adjusts the weights and biases accordingly. This process completes one epoch. Consequently, we can calculate the energy costs associated with forward and backward propagation, leading to an understanding of the overall energy consumption for training, model evaluation, and subsequent inference. In the following subsections, we showcase the computational complexity of training and evaluating the aforementioned \acp{nn}.

\subsection{Computational Complexity of MLP, CNN, KAN and transformer}
\paragraph{\acp{mlp}}
\label{sec:mlp}
Assuming an \ac{mlp} architecture with $L$ dense layers as depicted in Fig. \ref{fig:MLP_model}, where $L=K+2$, the total number of \ac{FLOPS} for the forward propagation of a single input sample can be expressed as: 
\begin{equation}
\small
\begin{aligned}
 M_{\mathrm{MLP}} &= \sum_{l=1}^{L-1} \Bigg( 
  \underbrace{(2   M_{l-1}   M_{l})}_{\text{Weight and bias}}
  + \underbrace{2   M_{l}}_{\text{Summation and activation}} \Bigg),
\end{aligned}
\label{eq:tot_mlp_fp}
\end{equation}
where the first term in the sum represents one multiplication and one addition ($=2$ \ac{FLOPS}) related to the weight and bias corresponding to the edges times the number of nodes in that layer ($=M_{l-1}   M_{l}$), while the second term represents the summation of all the values of the incoming edges and the application of the activation function ($=2$ \ac{FLOPS}) times the number of nodes in that layer ($M_{l}$). 

In general, however, training consists of multiple epochs and batches. Therefore, the total complexity of forward propagation is:

\begin{equation}
\begin{aligned}
        M_{\mathrm{MLP,FP}}&= N_{\mathrm{epochs}}  N_{\mathrm{batch}}  B   M_{\mathrm{MLP}}\\
    &=N_{\mathrm{epochs}}  N_{\mathrm{S,T}}  M_{\mathrm{MLP}}. 
    \label{eq:tot_mlp_epochs}
\end{aligned}
\end{equation}

\paragraph{\acp{cnn}}
Typical \acp{cnn} consist of three types of layers: dense layers, convolutional layers, and pooling layers.  For a convolutional layer, with input tensor of size $I_{\mathrm{r}}\times I_{\mathrm{c}}\times C_{\mathrm{in}}$, the corresponding computational complexity is delivered as:
\begin{equation}
\small
\begin{aligned}
\label{eq:tot_conv}
M_{\text{CONV}} =  \underbrace{\Bigg( \frac{I_{\text{r}} - K_{\text{r}} + 2P_{\text{r}}}{S_{\text{r}}} + 1\Bigg)}_{\text{Output height}}
  \underbrace{\Bigg( \frac{I_{\text{c}} - K_{\text{c}} + 2P_{\text{c}}}{S_{\text{c}}} + 1\Bigg)}_{\text{Output width}} \\
  \underbrace{\Bigg( C_{in}  K_{\text{r}}   K_{\text{c}} + 1\Bigg)}_{\text{Computation per filter}}   \underbrace{N_{\text{f}}}_{\text{No. of filters}},
\end{aligned}
\end{equation}
where $K_{\mathrm{r}}$ and $K_{\mathrm{c}}$ are the kernel (filter) height and width, $P_{\mathrm{r}}$ and $P_{\mathrm{c}}$ are the padding sizes for height and width, $S_{\mathrm{r}}$ and $S_{\mathrm{c}}$ are the stride values for height and width, and $N_{\mathrm{f}}$ is the number of filters in the layer. 
For down-sampling the input tensor data, a pooling layer is needed, and the number of \ac{FLOPS} of such layer is expressed as:
\begin{equation}
\begin{aligned}
\label{eq:tot_pool}
M_{\text{POOL}} = \underbrace{\Bigg(\frac{I_{\text{r}} - K_{\text{r}}}{S_{\text{r}}} + 1\Bigg)}_{\text{Output height}} 
  \underbrace{\Bigg(\frac{I_{\text{c}} - K_{\text{c}}}{S_{\text{c}}} + 1\Bigg)}_{\text{Output width}} 
  \underbrace{C_{\text{in}}}_{\text{Input channels}}.
\end{aligned}
\end{equation}

Thus, the total number of \ac{FLOPS} for the forward propagation of a single input sample in a \ac{cnn} can be calculated by summing the contributions across all layer types~\cite{bertalanivc2024carmel}.
As a result, the total \ac{FLOPS} of a \ac{cnn} is: 
\begin{equation} 
\begin{aligned}
    M_{\mathrm{CNN, FP}} &= N_{\mathrm{epochs}}  N_{\mathrm{S,T}}  \\
    &\left(\sum_{l=1}^{L_{c}} M_{\mathrm{CONV}} + \sum_{l=1}^{L_{p}} M_{\mathrm{POOL}} + M_{{\mathrm{MLP}}}\right).
    \label{eq:tot_cnn_fp} 
\end{aligned}
\end{equation}
\begin{figure*}[t]
   \centering
    \begin{subfigure}[b]{0.25\textwidth}
        \centering
        \includegraphics[width=\linewidth]{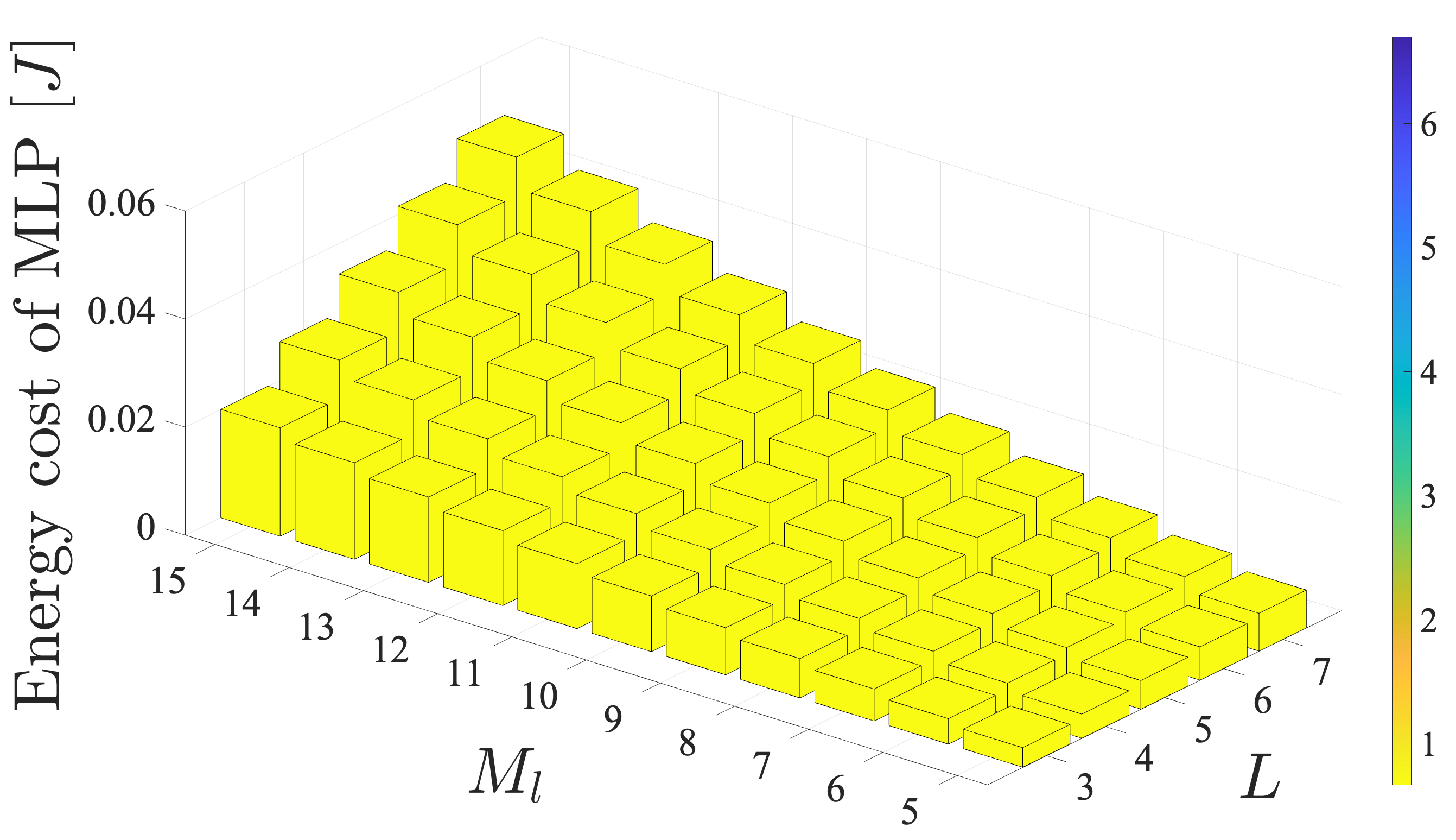}
        \caption{Energy cost of training a MLP versus $M_{l}$ and $L$}
        \label{fig:subfig1}
    \end{subfigure}%
    \hfill
    \begin{subfigure}[b]{0.25\textwidth}
        \centering
        \includegraphics[width=\linewidth]{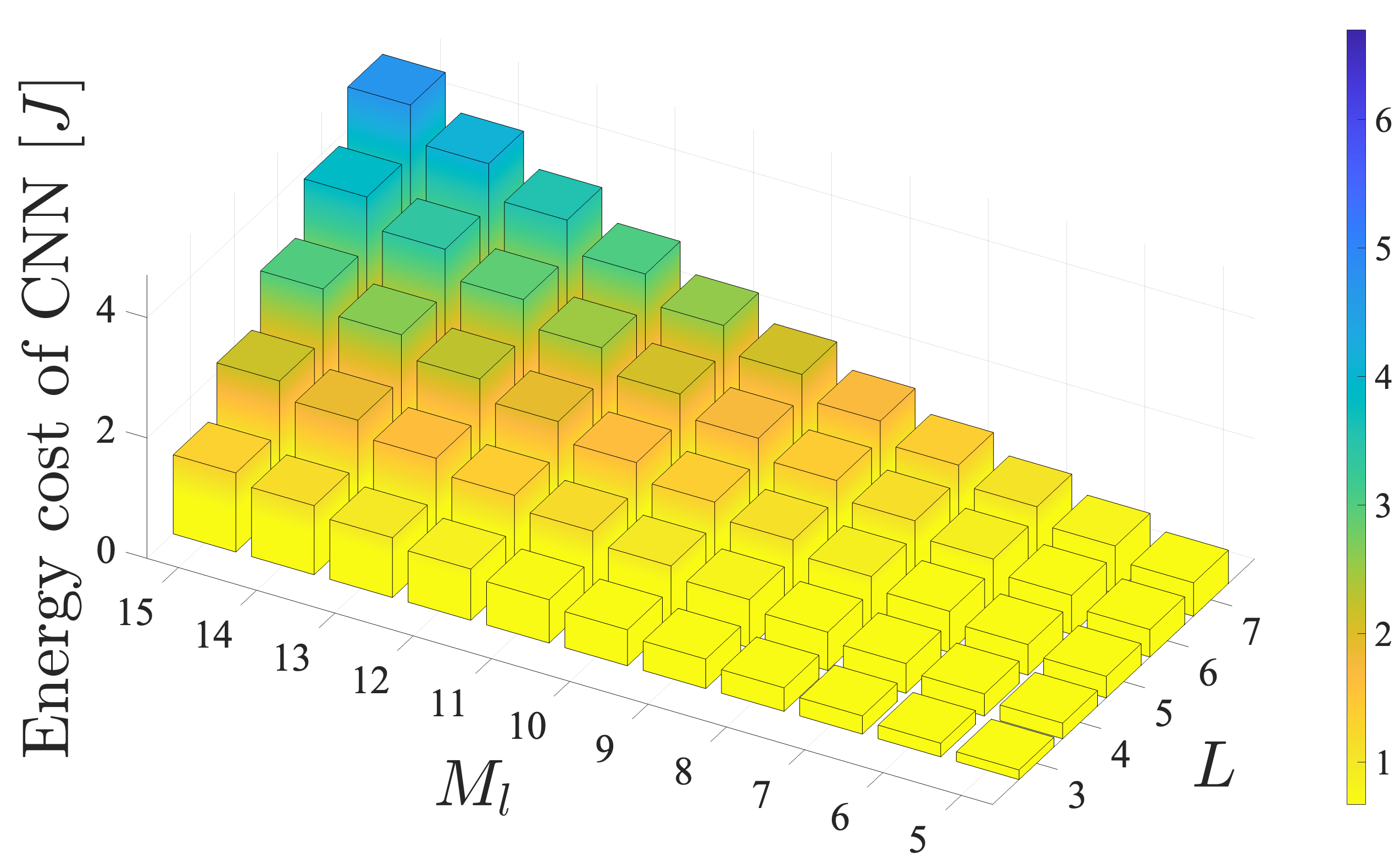}
        \caption{Energy cost of training a CNN versus $M_{l}$ and $L$}
        \label{fig:subfig2}
    \end{subfigure}%
    \hfill
    \begin{subfigure}[b]{0.25\textwidth}
        \centering
        \includegraphics[width=\linewidth]{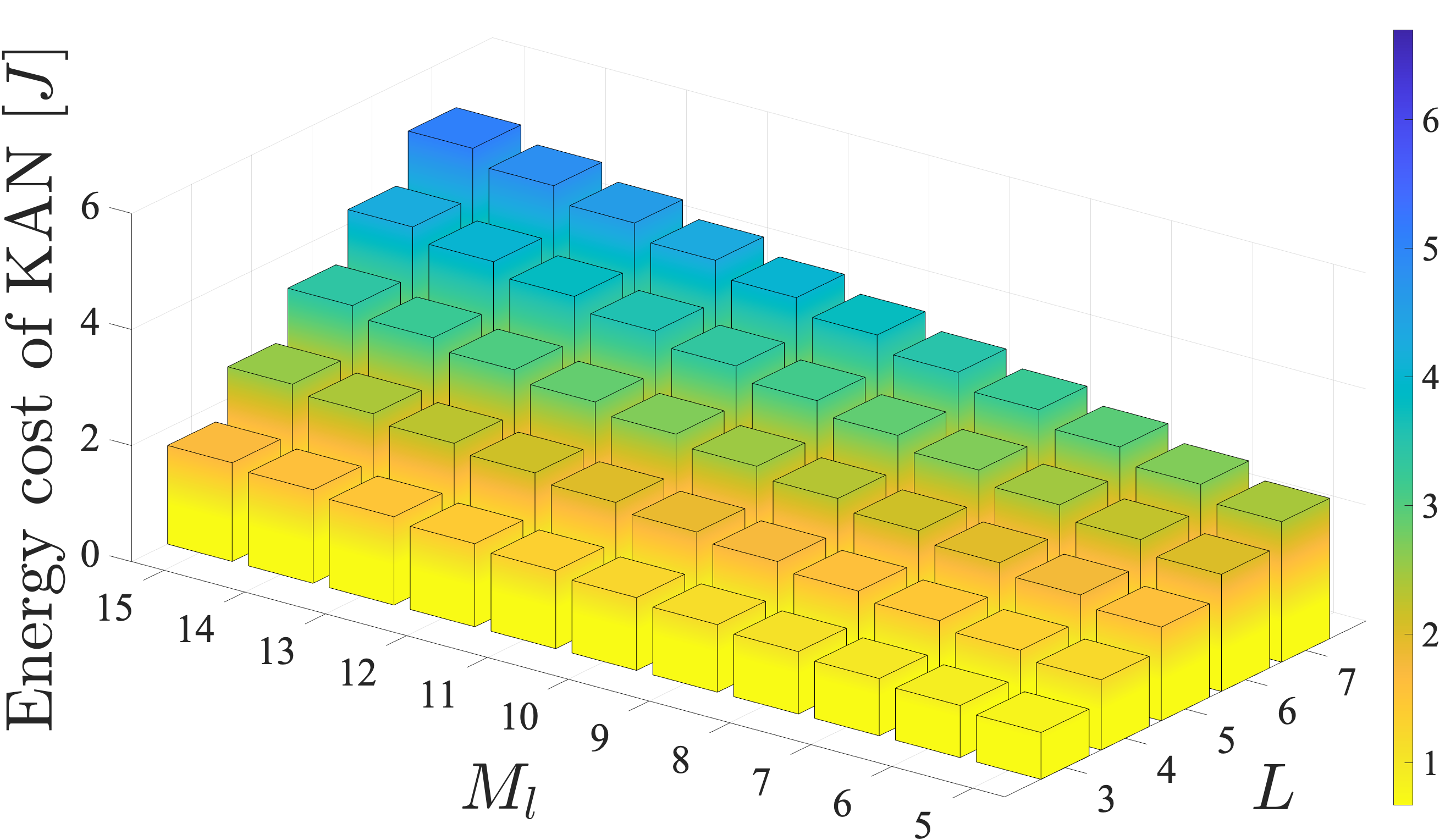}
        \caption{Energy cost of training a KAN versus $M_{l}$ and $L$}
        \label{fig:subfig3}
    \end{subfigure}%
    \hfill
    \begin{subfigure}[b]{0.25\textwidth}
        \centering
        \includegraphics[width=\linewidth]{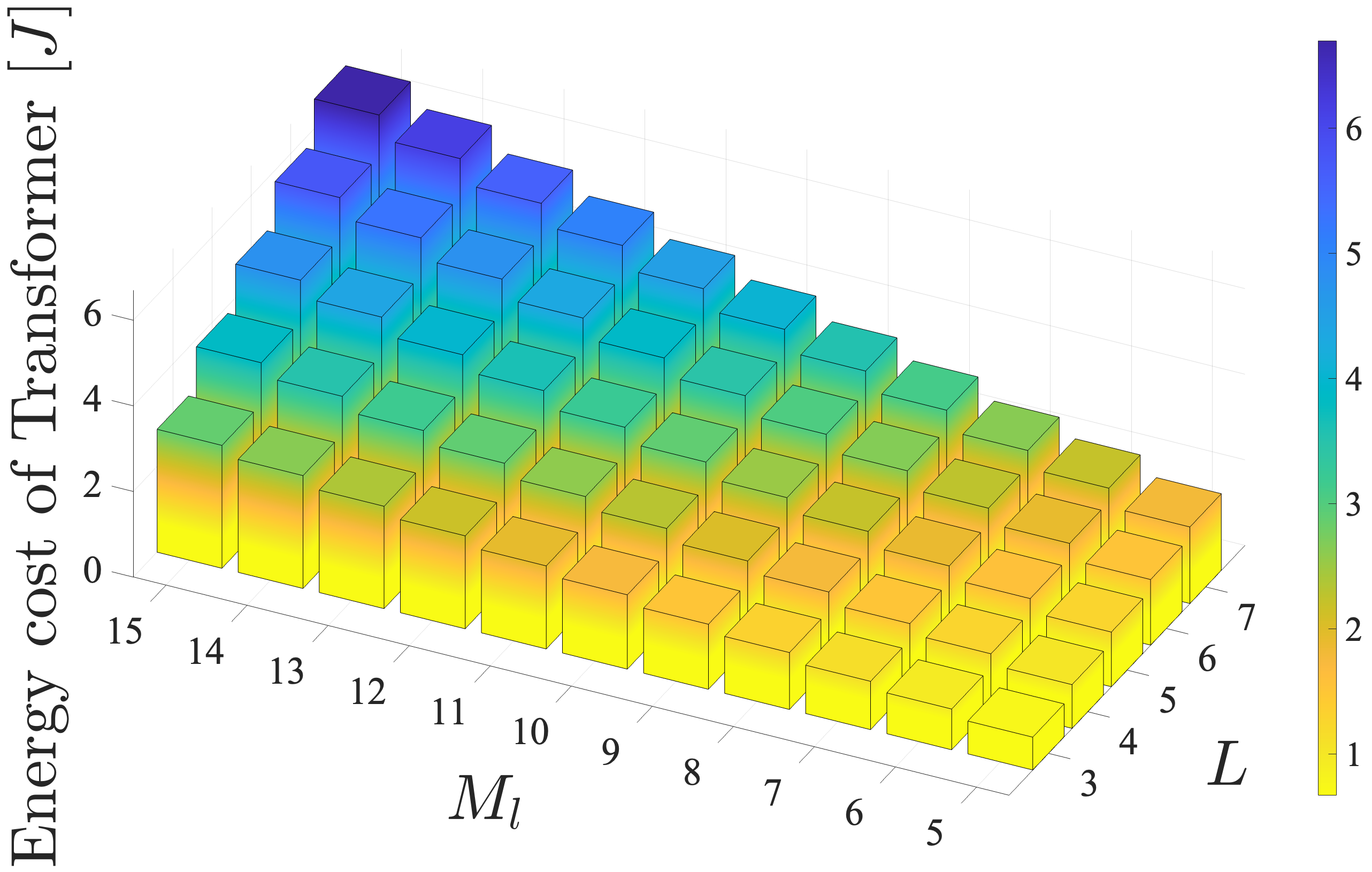}
        \caption{Energy cost of training a transformer versus $M_{l}$ and $L$}
        \label{fig:subfig4}
    \end{subfigure}

    \caption{Energy cost of training different models with respect to the number of nodes ($M_{l}$) and layers ($L$).}
    \label{fig:subfig_big}
    \vspace{-10pt}
\end{figure*}
\paragraph{\acp{kan}}
\acp{kan} leverage the Kolmogorov-Arnold representation theorem to decompose complex multivariate functions into univariate sub-functions. This is achieved through the use of learnable B-spline activation functions and shortcut paths~\cite{KAN_intro}. Unlike \acp{mlp}, which apply fixed activation functions to nodes, \acp{kan} feature learnable activation functions on edges, providing greater flexibility and interpretability. According to~\cite{kan_flops}, the computational complexity of a single \ac{kan} layer can be computed as: 
\begin{equation}
\small
\begin{aligned}
      M&_{\mathrm{KAN}}= 
      \sum_{l=1}^{L-1} (
      \underbrace{M_{\mathrm{NLF}}   M_{l-1}}_{\text{From B-spline}}\\
      &+\underbrace{\left(M_{l-1}   M_{l}\right)}_{\text{Input and output}}    \underbrace{\left[9K  \left(G+1.5K\right)+2G-2.5K+3\right])}_{\text{B-spline and grid}}\\
      &= \sum_{l=1}^{L-1} ( M_{\mathrm{NLF}}  M_{l-1}+\left(M_{l-1}  M_{l}\right)  M_{\mathrm{B}}),
      \label{eq:KAN_layer}
\end{aligned}
\end{equation}
where $M_{\mathrm{NLF}}$ denote as the number of \ac{FLOPS} contributed from the B-spline activation function across all input elements. The second term accounts for the computational costs arising from the combination of input and output dimensions ($M_{l-1}$ and $M_{l}$) with the operations performed by the B-spline transformation. These operations include the iterative evaluation of spline basis functions and the associated transformations, governed by the spline order ($K$) and the number of intervals in the grid ($G$). The total \ac{FLOPS} of a \acp{kan} network is calculated as the sum of all \ac{kan} layers similarly as in the case of \acp{mlp} in Eq.~(\ref{eq:tot_mlp_epochs}), where total $L$ layers can be expressed as:
\begin{equation}
\begin{aligned}
    M_{\mathrm{KAN,FP}}=&N_{\mathrm{epochs}}  N_{\mathrm{S,T}}  M_{\text{KAN}}.
    \label{eq:tot_kan_fp} 
\end{aligned}
\end{equation}

\paragraph{Transformer} Based on the attention mechanism, the transformers \cite{vaswani2017attention} represent a paradigm shift in deep learning, replacing traditional architectures such as \ac{cnn} with self-attention mechanisms. While enabling significant breakthroughs in \ac{ai} through efficient parallel processing and robust modeling of long-range dependencies, they are relatively complex. In addition to attention layers, the transformer architecture also includes \ac{mlp}, normalization, and positional encoding, making the analytic expression for the FLOP computation intractable. Following existing literature~\cite{ouyang2023understanding} and approximations\footnote{\url{https://www.gaohongnan.com/playbook/training/how_to_calculate_flops_in_transformer_based_models.html}}, we approach the estimation as follows:

\begin{equation}
\small
\begin{aligned}
\label{eq:tot_att}
M_{\text{ATT}} =  2    (\underbrace{C   N_{\text{embed}}   3    N_{\text{embed}}}_{\text{K, Q, V positional embedding}}
+ \underbrace{C^{2}   N_{\text{embed}}}_{\text{Attention scores}} \\
+ \underbrace{N_{\text{head}}    C^{2}   N_{\text{embed}} / N_{\text{head}}}_{\text{reduce}} +  \underbrace{C   N_{\text{embed}}   N_{\text{embed}}}_{\text{Projection}}),
\end{aligned}
\end{equation}

\noindent where $C$ is the context length, $N_{\text{embed}}$ is the size of the embedding and $N_{\text{head}}$ is the number of heads. The complexity of the decoder in a transformer is then given by: 
\begin{equation}
\small
\begin{aligned}
\label{eq:tot_tr}
M_{\text{TR}} =  N_{\text{decoder\_blocks}}   (
M_{\text{ATT}}
+ \underbrace{4   C   N_{\text{embed}}   FFS}_{\text{MLP blocks}}),
\end{aligned}
\end{equation}
where $N_{\text{decoder\_blocks}}$ and $FFS$ stand for the number of decoder blocks and the feed-forward size, respectively. The total complexity of forward propagation is:
\begin{equation}
\begin{aligned}
    M_{\mathrm{TR,FP}}
    =&N_{\mathrm{epochs}}  N_{\mathrm{S,T}}  M_{\text{TR}}.
    \label{eq:tot_tr_fp} 
\end{aligned}
\end{equation}
For more specific cases, such as GPT-like architectures, a final dense layer is also needed. However, this work focuses on a more general formulation of transformer models.
\subsection{\textcolor{black}{Energy Estimation for the Model Training}}
To determine the computational complexity of training a model, backward propagation needs to also be considered. The standard approximation, also confirmed in~\cite{BPis2times}, is that the backward propagation requires two times the \ac{flops} of the forward propagation. In particular, backward propagation requires matrix multiplications for updating the weights and propagating the gradient. Thus, we can approximate the computing complexity of training a model with Eq. \ref{eq:tot_mlp} and obtain the energy consumption of the entire training process with E1. \ref{eq:E_mlp}. 
\begin{equation}  M_{\mathrm{model,tot}}\approx  3M_{\mathrm{model,FP}}.
\label{eq:tot_mlp}
\end{equation}
\begin{equation}
E_{\mathrm{train}}
= \frac{3M_{\mathrm{model,FP}} [\mathrm{FLOPs}]}{PU_{\mathrm{performance}} [\mathrm{FLOPs/s/W]}}. 
\label{eq:E_mlp}
\end{equation}
From Eqs.~(\ref{eq:tot_mlp_epochs}), (\ref{eq:tot_cnn_fp}), (\ref{eq:tot_kan_fp}), (\ref{eq:tot_tr_fp}), and~(\ref{eq:E_mlp}), we can observe that the training data set size scales linearly with energy consumption of the model training, as it does not affect the cost of a single forward/backward pass. On the other hand, training the same model on a state-of-the-art GPU with a processing power of $312~[TFLOPs/s]$ and power consumption of $400~[W]$ is approximately $28$ times more energy efficient than using a CPU, which operates at $13.824~[TFLOPs/s]$ and consumes $500~[W]$.
\paragraph*{\textcolor{black}{Numerical and Empirical Validation}} Fig.~\ref{fig:subfig_big} illustrates the relationship between the energy consumption of a model and two architectural parameters: the number of layers and the number of nodes in a layer. As the architectures of the models become deeper and wider, the computational complexity also increases, and this leads to higher energy consumption. For instance, in the case where $M_{l}=10$ and $L=5$, the energy consumption of the transformer, \ac{kan}, and \ac{cnn} models are approximately $173$, $149$, and $81$ times greater than that of the \ac{mlp} model. However, some architectures, such as KANs, are known to perform better with fewer layers ~\cite{KAN_intro}, therefore it is unlikely to find a KAN with $L>5$. While at a first glance, it seems that MLPs may be significantly more energy efficient than KANs, the actual difference for the same performance will be less prominent for many applications. CNNs, and especially transformers are known to be computationally demanding and subsequently exhibit high energy consumption, while at the same time larger architectures proved better performance in various application areas.
\begin{figure}[!htb]
    \centering
    \includegraphics[width=0.48\textwidth]{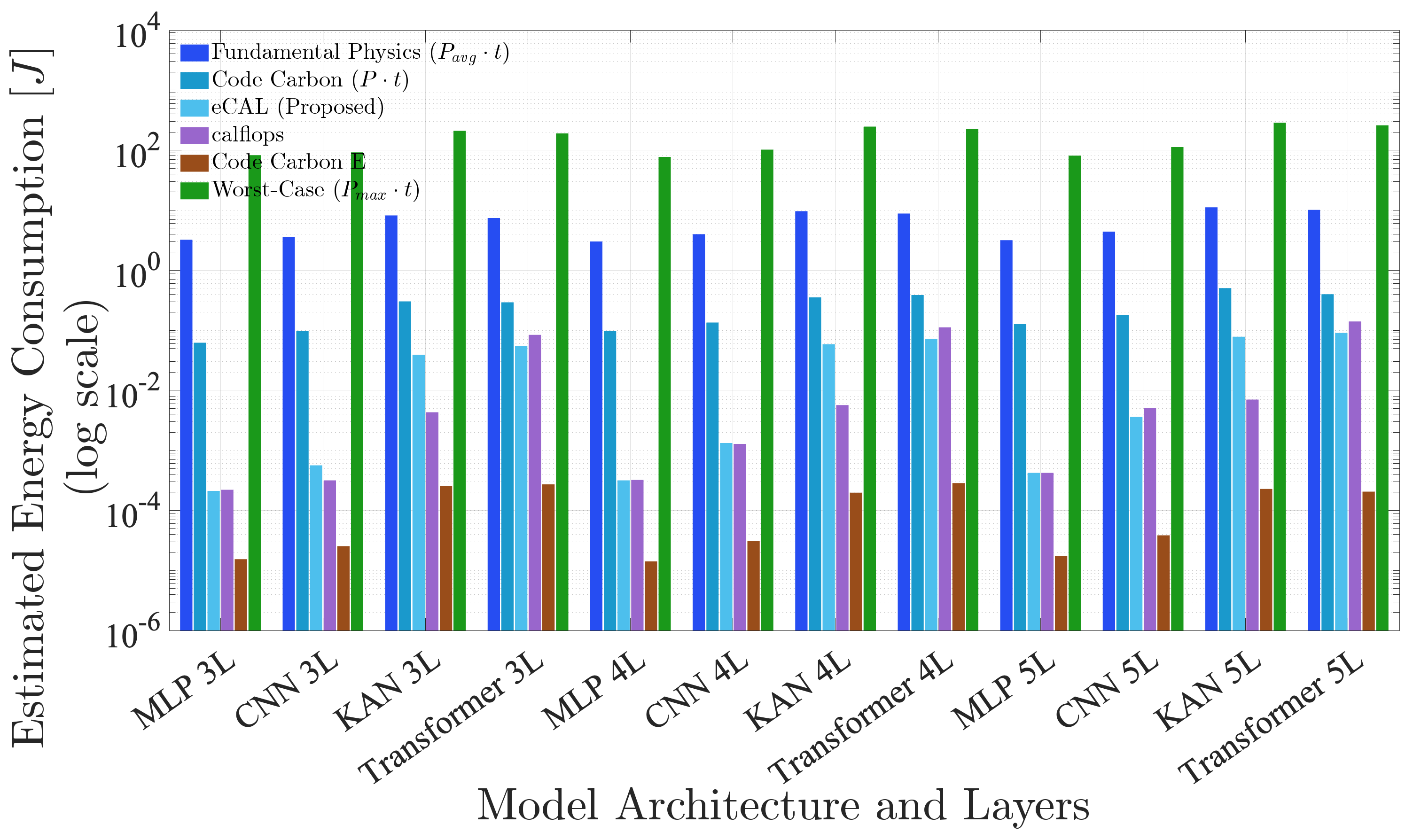}
    \caption{Training energy consumption benchmarks for the selected  fundamental neural architectures.}
    \label{fig:CompModelsvsMeasurements}
    \vspace{-15pt}
\end{figure}

\textcolor{black}{Fig. \ref{fig:CompModelsvsMeasurements} benchmarks the model training energy required by eCAL against (a) a simple fundamental physics approach $E = P_{avg} \cdot t$ that considers the average power consumption of the GPUs over the training time, (b) the open-source calflops\footnote{calflops,https://pypi.org/project/calflops/0.0.2/} library able to compute the theoretical amount of FLOPs  for neural networks implemented in pytorch, (c) Code Carbon\footnote{Code Carbon, https://github.com/mlco2/codecarbon}, and (d) a worst case scenario that assumes the GPU is used at maximum capacity during the entire training time computed as $E = P_{max} \cdot t$. For eCAL as per Eq.~(\ref{eq:E_mlp}),  calflops and worst case, we set the value of $P_{max} = 10 W$ as typical for an Apple M2 GPU. For the fundamental physics we measured the average power used by the GPU at $P_{avg} = 0.38 W$\footnote{This study can be replicated on other set-ups with https://github.com/sensorlab/eCAL/blob/main/scripts/
All\_Theoretical\_Empirical\_Training\_Energy.py} using the PowerMetrics tool. Code Carbon outputs energy, power and time duration measurements. It measures the power and time duration, plotted as Code Carbon ($P \cdot t$), through system tools, such as PowerMetrics for Apple silicon or a dedicated python library for NVIDIA GPUs, while it computes the energy, plotted as Code Carbon E, based on a model. The results show that eCAL and calflops provide very similar energy footprint estimated for MLPs, CNN and Transformers based networks. For MLPs, the difference is negligible, for CNNs eCAL slightly overestimates compared to calflops, while for transformers it slightly underestimates. The slight differences are due to the fact that it is difficult to harmonize all hyperparameters between the theoretical expressions that reflect the neural network architectures and corresponding software implementation that instantiates pytorch models that calfplos relies on. For KANs as newer architectures, calflops underestimates compared to the eCAL due to the fact that the implementation of the architecture relies on several optimizations that reduce the number of FLOPs compared to the theoretical expressions.}

\textcolor{black}{The results also show that assuming average power usage during the entire training cycle with the fundamental physics expression $E = P_{avg} \cdot t$ overestimates the consumption. The overestimation is more prominent in simple examples used for this study where the data loading and model saving during training take a significant part of the training time when the GPU is not active. However, when these steps become negligible, this estimation will close in to eCAL and calflops showing the validity of our approach. Code Carbon E measures significantly less consumed energy while Code Carbon ($P \cdot t$) measures significantly more than eCAL and calflops. These results also confirm the existing challenges regarding measuring energy and CO$_{2}$ footprint of software \cite{pathania2025calculating} and confirm the timeliness of our work. }

\subsection{Model Evaluation}
The model evaluation starts once training is complete. This functionality tests the performance of the model on a separate dataset. During the evaluation, the model processes the data using forward propagation without any adjustments to its parameters. Therefore, the energy consumption of model evaluation is expressed as:
\begin{equation}
    {E_{\mathrm{eval}}
    = \frac{M_{\mathrm{model}}  N_{\mathrm{S,E}}}{PU_{\mathrm{performance}}}}.
    \label{eq:E_eval}
\end{equation}

\noindent It is worth noting that the total energy consumption during training and evaluation is significantly influenced by the chosen evaluation strategy. While a simple train-test split requires training the model only once, more robust techniques like $k$-fold cross-validation necessitate $k$ complete training cycles. For $k$-fold cross-validation, the total training energy consumption becomes $k$ times the value given in Eq.~(\ref{eq:E_mlp}), as the model must be retrained from scratch for each fold. Similarly, techniques such as nested cross-validation or repeated $k$-fold cross-validation further multiply the required training computations and corresponding energy costs, in which we have also incorporated $k$-fold cross-validation considerations into eCAL calculator.
\begin{figure}[t]
    \centering
    \includegraphics[width=0.48\textwidth]{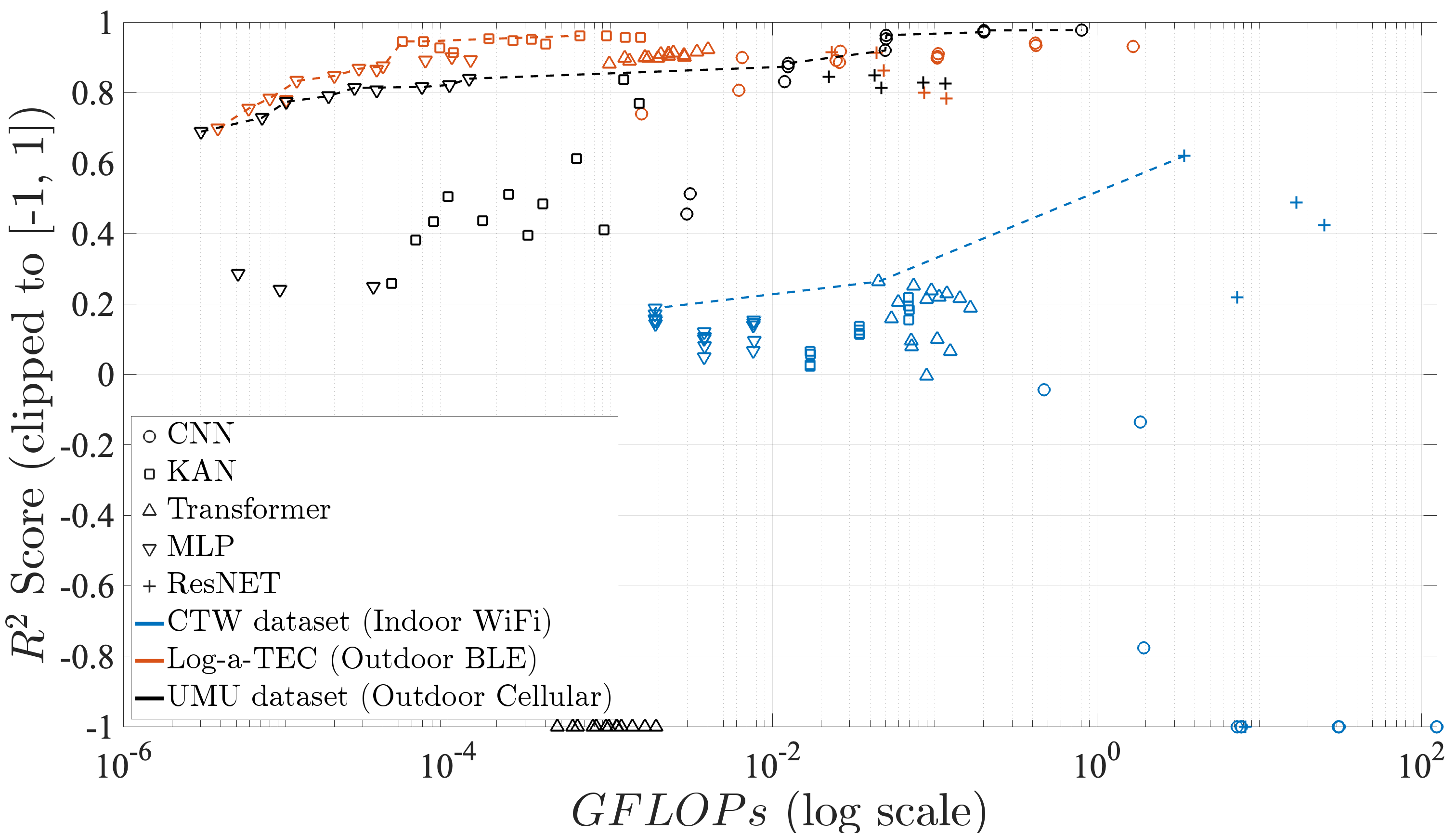}
    \caption{Model performance (higher $R^2$
   is better) vs. computational cost (GFLOPs) for five different architectures across the three datasets. The dashed lines indicate the Pareto front.}
    \label{fig:paretofront}
    \vspace{-15pt}
\end{figure}
\begin{table*}[t]
\centering
\caption{Total energy consumption of developing ($E_{\mathrm{D}}$) for different models, with contributions of individual data-manipulating components.}
\label{tab:energy_comparison}
\footnotesize
\begin{tabular}{lcccc}
\hline
\textbf{Component} & \textbf{MLP} & \textbf{CNN} & \textbf{KAN} & \textbf{Transformer} \\
\hline
$E_{\mathrm{DC}}~[J]$ & $8.44 \times 10^{-2}$ ($89.12\%$) & $8.44 \times 10^{-2}$ ($12.37\%$) & $8.44 \times 10^{-2}$ ($6.31\%$) & $8.44 \times 10^{-2}$ ($3.72\%$) \\
$E_{\mathrm{pre}}~[J]$ & $1.54 \times 10^{-6}$ ($0.00\%$) & $1.54 \times 10^{-6}$ ($0.00\%$) & $1.54 \times 10^{-6}$ ($0.00\%$) & $1.54 \times 10^{-6}$ ($0.00\%$) \\
$E_{\mathrm{train}}~[J]$ & $1.03 \times 10^{-2}$ ($10.86\%$) & $5.97 \times 10^{-1}$ ($87.48\%$) & $1.25$ ($93.53\%$) & $2.18$ ($96.12\%$) \\
$E_{\mathrm{eval}}~[J]$ & $1.72 \times 10^{-5}$ ($0.02\%$) & $9.95 \times 10^{-4}$ ($0.15\%$) & $2.08 \times 10^{-3}$ ($0.16\%$) & $3.64 \times 10^{-3}$ ($0.16\%$) \\
\hline
$E_{\mathrm{D}}~[J]$ & \textbf{$9.47 \times 10^{-2}$} & \textbf{$6.83 \times 10^{-1}$} & \textbf{$1.34$} & \textbf{$2.27$} \\
\hline
\end{tabular}%
\vspace{-15pt}
\end{table*}

\textcolor{black}{To illustrate the trade-off between model performance and computational complexity, and to demonstrate how eCAL supports Pareto-front analyses, we evaluate three wireless datasets (Wi-Fi, Bluetooth Low Energy (BLE), and cellular), four fundamental architectures from Section \ref{sec:e_train}, and a ResNet combining CNN and MLP blocks—a configuration known to perform well for wireless localization \cite{pirnat2022towards}. Fig.~\ref{fig:paretofront} shows model performance ($R^{2}$) versus computational complexity at inference time $M_{\mathrm{model,FP}}$.
For the univariate BLE dataset, the highest $R^{2}$ scores are achieved by MLP and KAN models forming the Pareto front, while CNN, ResNet, and Transformer models underperform with higher computational costs. Similar trends appear for the 1D multivariate UMU dataset, where the Transformer performs poorly ($R^{2}=-1$) and is notably outperformed by simpler MLPs. With more signals (multivariate data), CNN performance improves, and these models appear on the Pareto front. On the larger, more complex multivariate CTW dataset, which includes 2D antenna array shapes, Transformer-based models perform better and also lie on the Pareto front.
Overall, these results show that more complex neural architectures do not necessarily yield better performance. The balance between model complexity and accuracy depends on both the architecture and the nature and quality of the data.}
\section{Energy Cost of Inference}
\label{sec:E_inf}
The complexity of making an inference depends on the number of \ac{FLOPS} required for forward propagating with input sample size $N_{\mathrm{I,P}}$, i.e., $N_{\mathrm{inf}}=M_{\mathrm{model}}  N_{\mathrm{I,P}}$. For example, when we adopt the aforementioned MLP model with $N_{\mathrm{I,P}}=51$, the complexity is calculated as $N_{\mathrm{inf}}=236 \times 51 = 12,036 $ \ac{FLOPS}. The energy consumption of the forward propagation and the corresponding inference can be calculated as: 
\begin{equation}
\begin{aligned}
    E_{\mathrm{inf}}
    = \frac{M_{\mathrm{model}}  N_{\mathrm{I,P}}}{PU_{\mathrm{performance}}}. 
\end{aligned}
\label{eq:E_pred}
\end{equation}

We can then observe from Eqs.~(\ref{eq:E_eval}) and~(\ref{eq:E_pred}) that the cost of inference is the same as in evaluation when $N_{\mathrm{S,E}}$ is equal to $N_{\mathrm{I,P}}$. Thus, we can conclude that the computational complexity between finishing one training and inference is on the magnitude of $3  N_{\mathrm{epochs}}  N_{\mathrm{S,T}}/N_{\mathrm{I,P}}$. This indicates that the training process involves significantly more computational operations, particularly when the number of epochs or 
$N_{\mathrm{S,T}}>>N_{\mathrm{I,P}}$. On the other hand, the energy consumption per bit depends on the model complexity and hardware capability. Moreover, the factor of three in the energy consumption per bit during training versus inference highlights the additional computational load inherent to training.
\section{eCAL: the Energy Cost of AI Lifecycle}
\label{sec:e2e}
In this section, we describe the end-to-end energy consumption of the AI model lifecycle for making inferences with the aforementioned neural network architectures. First, we consider the energy consumption involved in developing the model, as illustrated on the left in Fig.~\ref{fig:system_model}, which can be expressed as in Eq. (\ref{eq:E_aiot_e2e_deploy}) with the energy consumption per bit expressed in Eq. (\ref{eq:E_aiot_e2e_deploy_perbit}).
\begin{equation}
E_{\mathrm{D}}=E_{\mathrm{DC}}+E_{\mathrm{pre}}+E_{\mathrm{train}}+E_{\mathrm{eval}}.\label{eq:E_aiot_e2e_deploy}
\end{equation}
\begin{equation}
E_{\mathrm{D,b}}=\frac{E_{\mathrm{D}}}{f N_{\mathrm{S}}I_{\mathrm{S}}}.\label{eq:E_aiot_e2e_deploy_perbit}
\end{equation}

To compare the energy consumption of all the data manipulation components required for developing and deploying an AI-enhanced system, we assume that the models are trained and evaluated using 5-fold cross-validation (as discussed in Section \ref{sec:e_train}). The models require $256$ samples with an input size of $10$, i.e., $N_{\mathrm{S}}  I_{\mathrm{S}}= 2560$, collected from a  device over a wireless network configured with the following parameters: $P_{\mathrm{T}}=200~[mW]$, $P_{\mathrm{R}}=5~ [mW]$, $R_{\mathrm{T}}=R_{\mathrm{R}}=10~[Mbps]$, $N_{\mathrm{\_, cycle},l}=100~[c/b]$, and $P_{\mathrm{\_,cycle}}=10^{-10}~[W/c]$. We also assume the link between the  device and the server is stable, and there is no retransmission needed. Each layer introduces \ac{dp} and \ac{cp} overhead of $10\%$ and $5\%$, respectively. Furthermore, normalization is utilized as the data transformation procedure, as discussed in Section \ref{sec:preproc}. Finally, model training requires $10$ epochs ($N_{\mathrm{epochs}}=10$) and utilizes $5$-fold validation. The energy consumption of this example is shown in Table~\ref{tab:energy_comparison} and reveals the contribution of each component to the total energy cost of developing models. Specifically, data collection dominates the total energy cost for developing a very simple model based on a 3-layer, 10-nodes-per-layer \ac{mlp} architecture. For the same parameters, the relative contribution of the data collection decreases for the KAN, CNN and transformer architectures, making the energy cost of training the primary contributor. The eCAL calculator enables changing the hyperparameters of the four architectures while also adding well known CNN architectures such as ResNets~\cite{he2016residual}, VGGs~\cite{simonyan2014vgg} or transformer-based such as Baichuan 2~\cite{yang2023baichuan}.

\begin{figure}[t]
    \centering
    \includegraphics[width=0.48\textwidth]{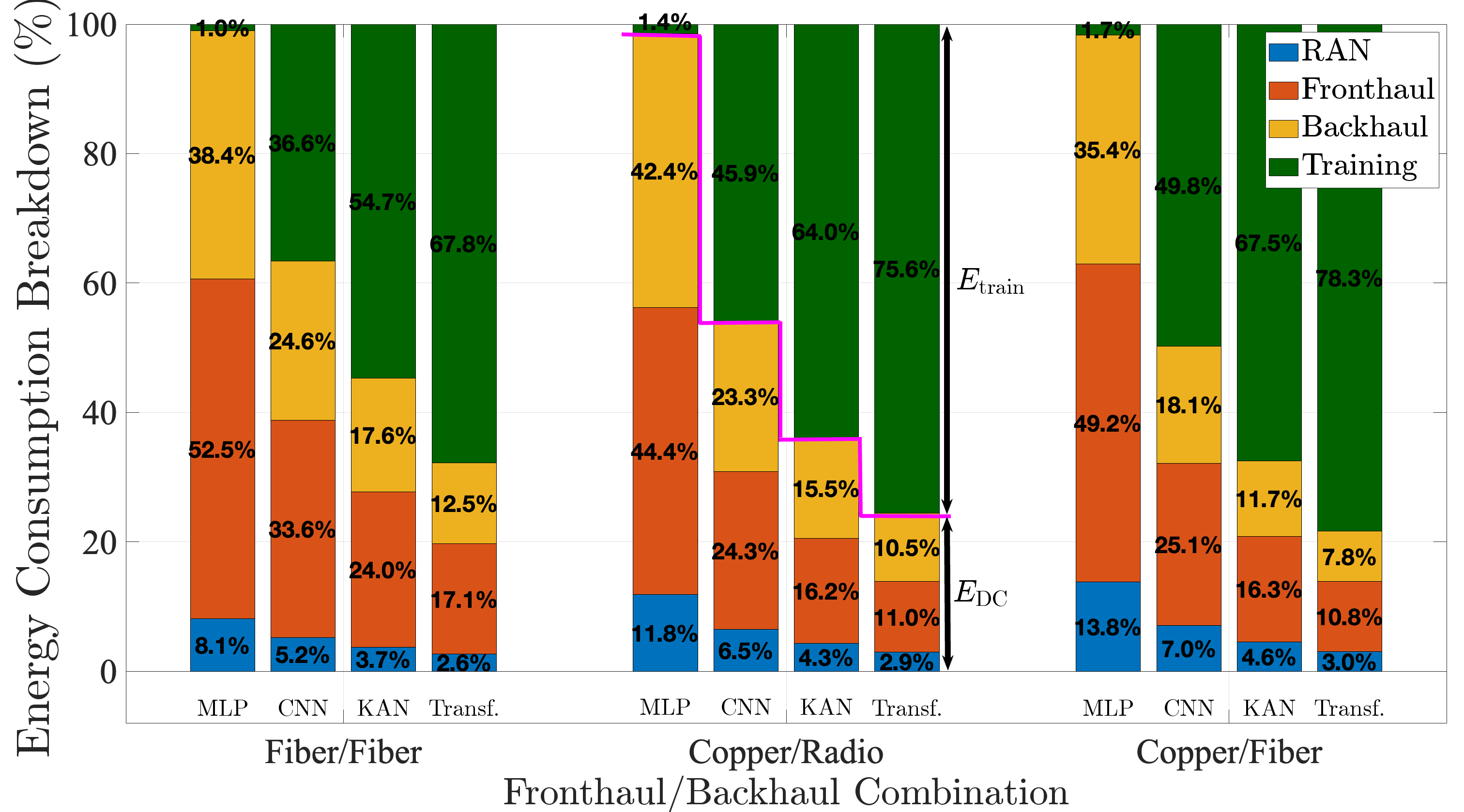}
    \caption{Comparison of energy consumption of training ($E_{\mathrm{train}}$, highlighted in upper part of the bars) and data collection ($E_{\mathrm{DC}}$, highlighted in lower part of the bars, divided into three parts: RAN, Fronthaul, and Backhaul for different models in different Fronthaul/Backhaul combinations.}
    \label{fig:wired_figure}
    \vspace{-10pt}
\end{figure}

\textcolor{black}{To complement Table~\ref{tab:energy_comparison} and provide a more comprehensive energy profile beyond the analysis in Section~\ref{sec:E_T}, Fig.~\ref{fig:wired_figure} shows the energy consumption breakdown that also includes wired portions of the network including fronthaul and backhaul, by using the microcell splits reported in~\cite{PolimiEuropacable2021}. We examine three fronthaul/backhaul combinations, i.e. Fiber/Fiber, Copper/Radio, and Copper/Fiber to illustrate how the energy mix shifts when wired paths dominate. For computationally light models such as a relatively shallow \ac{mlp}, deployment energy is no longer dominated by \ac{ran}-based data collection as in Table~\ref{tab:energy_comparison}; the wired infrastructure becomes the primary contributor, reaching roughly 90\% of the total in the Fiber/Fiber case. For complex models such as the simple transformer considered in this work, training remains the largest single component in energy consumption, but its share drops markedly once wired costs are included, which falling from over 96\% in the wireless-only view from Table~\ref{tab:energy_comparison} to approximately 68–78\% across the three scenarios as depicted in  Fig.~\ref{fig:wired_figure}. Importantly, the underlying absolute energies (in Joules) are unchanged: we add fronthaul and backhaul (derived from the microcell splits) to the per-model totals and then recompute percentage shares. The \ac{ran} data-collection energy is exactly the same as in Table~\ref{tab:energy_comparison}. Taken together, these results show a holistic energy assessment in modern hybrid networks, and emphasize the flexibility of the eCAL framework in modeling diverse scenarios.} 

Once the model is developed, it is packaged and deployed in the communication system, where it can produce inference in the form of continuous or discrete outputs. During its operational lifecycle, the model will be presented with input data and requested to produce the corresponding inference. Thus, we can express the associated energy cost as follows:
\begin{equation}
E_{\mathrm{inf,p}}=E_{\mathrm{DC}}+E_{\mathrm{pre}}+E_{\mathrm{inf}}.
\label{eq:E_inf_process}
\end{equation}

This means that we consider the energy cost of inference before the current model needs to be updated. The corresponding energy per bit of inference, once deployed, is calculated as follows: 
\begin{equation}
E_{\mathrm{inf,p,b}}=\frac{E_{\mathrm{inf,p}}}{fN_{\mathrm{I,P}}I_{\mathrm{S}}}.
\label{eq:E_inf_p_b}
\end{equation}

Next, we calculate the total energy cost over the lifetime of an AI model in the system,  also referred to as $eCAL_{\mathrm{abs}}$ (measured in [$J$]). It is defined as the sum of energy consumed during the model development, $E_{\mathrm{D}}$ and the energy required for performing an inference $E_{\mathrm{inf,p}}$, multiplied by the number of times $\gamma \in \left[0, \infty\right)$ the inference is performed with the currently deployed model. In addition, as edge cloud systems may use virtualization in developing and operating the models, we enable considering the impact of such technologies through an additional factor. As a result, $eCAL_{\mathrm{abs}}$ can be expressed as: 
\begin{equation}
eCAL_{\mathrm{abs}}=\left(1+\gamma_{\mathrm{v}}\right)\left(E_{\mathrm{D}}+\gamma E_{\mathrm{inf,p}}\right),
\label{eq:E_inf_tot}
\end{equation}

\noindent where $\gamma_{\mathrm{v}}$ represents the contribution from the virtualization, \textcolor{black}{with $\gamma_{\mathrm{v}} = 0$ corresponding to no virtualization. The reason for simplifying the estimation of the overhead introduced by the  virtualization technologies in eCAL through a factor $\gamma_{\mathrm{v}}$ is that the virtualization technologies in cloud systems are relatively well established while in communication systems they are still emerging. Furthermore, they depend on the configuration of the cloud-edge virtualization (e.g. bare-metal vs containers) and platform stacks (e.g. K8s vs K3s vs MicroK8s) as well as MLOps technology stack selection. For instance, there is significant difference in control-plane overhead on the platform layer of the CCRA depending on the number of control plane and worker plane nodes across K8s versions as shown in \cite{MUSIC2024619}. Furthermore, there is significant overhead difference between the standard O-RAN MLOps stack vs. NAOMI, a recently introduced distributed alternative \cite{COP2025104180}.  In the future, as technologies will mature and consolidate, it is  possible to set this factor to 0, tune the overhead in Eqs.~(\ref{eq:bits}) and incorporate compute and control plane overhead in Eq. (\ref{eq:E_inf_tot}) to enable more fine grained analytical study of virtualization impact.}
\begin{figure}[t]
    \centering
    \includegraphics[width=0.48\textwidth]{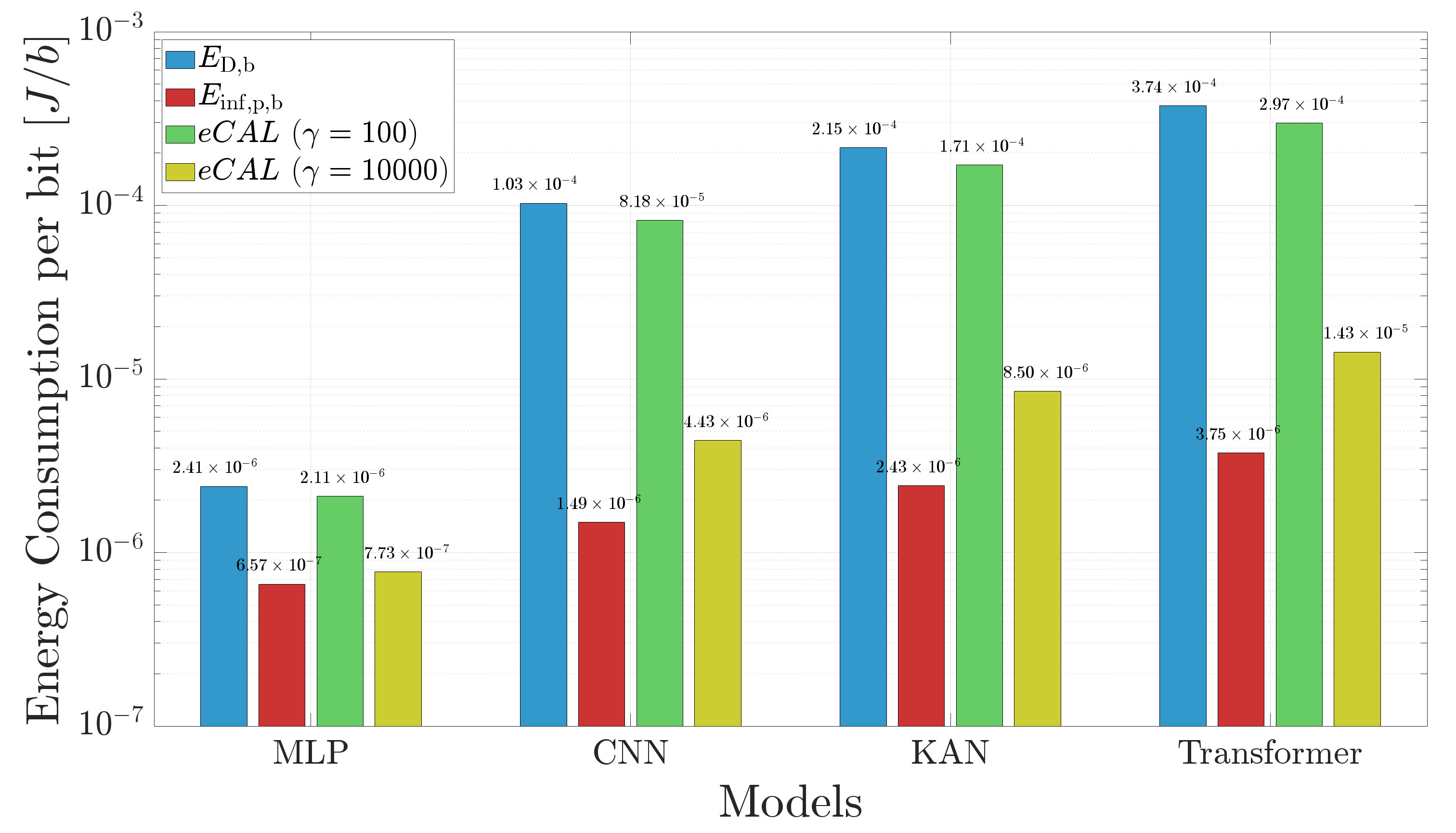}
    \caption{Comparison of energy consumption per bit in different data manipulation components over the lifecycle of the AI model.}
    \label{fig:E_inf_b}
    \vspace{-15pt}
\end{figure}

Finally, the proposed eCAL metric computing the corresponding energy consumption per bit over the lifecycle of the  currently deployed AI model can be expressed as:
\begin{align}
    eCAL=\frac{eCAL_{\mathrm{abs}}}{fI_{\mathrm{S}} (N_{\mathrm{S}}+\gamma N_{\mathrm{I,P})}}.\label{eq:e_inf_tot_b}
\end{align}
\begin{figure}[t]
    \centering
    \includegraphics[width=0.47\textwidth]{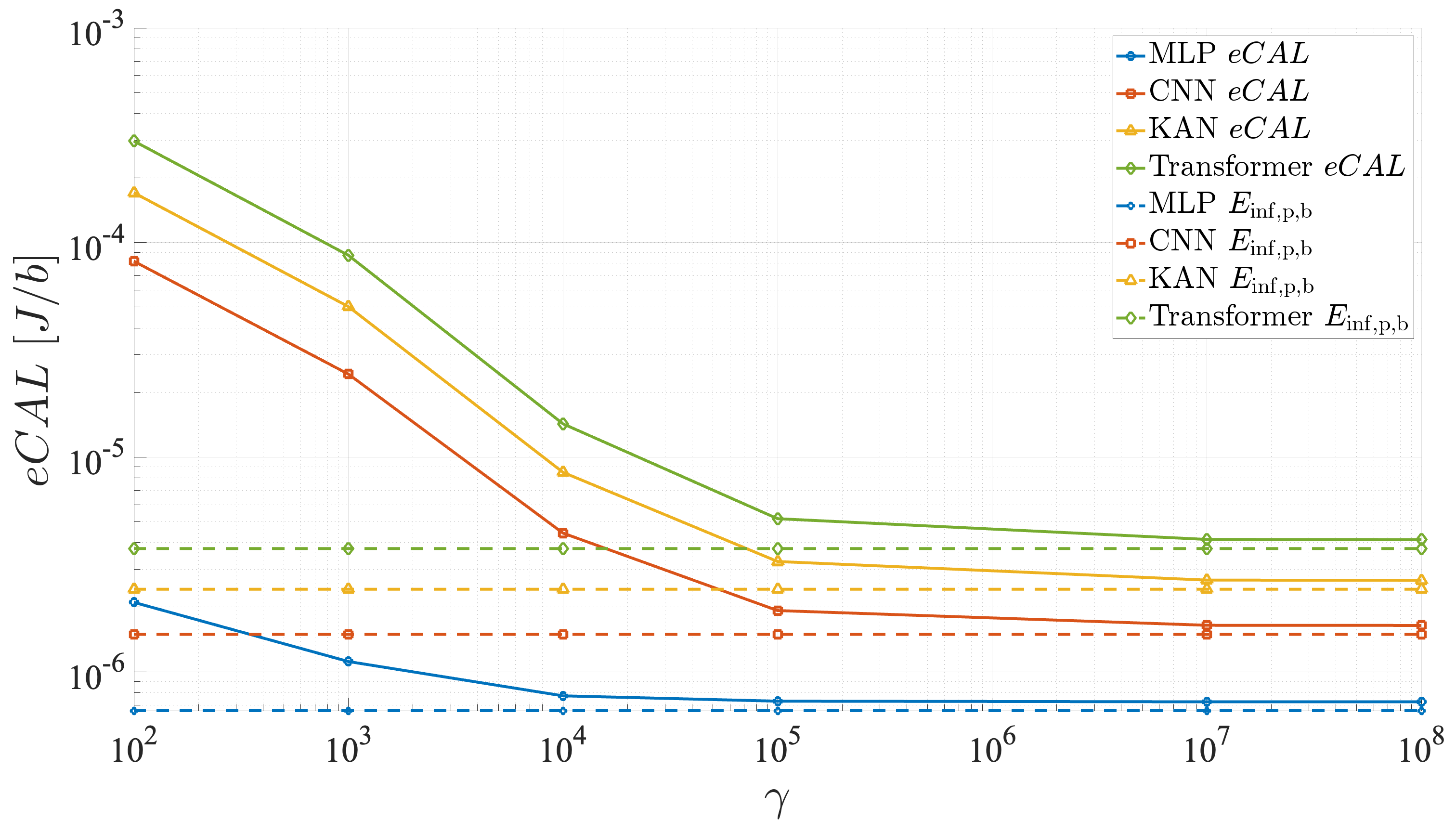}
    \caption{Energy cost of AI model  lifecycle ($eCAL$) for different models over the number of inferences ($\gamma$), log scale on both axes.}
    \label{fig:Ecal}
    \vspace{-15pt}
\end{figure}
\noindent To further quantify this relationship, we first compare the outcomes of Eqs.~(\ref{eq:E_aiot_e2e_deploy_perbit}), (\ref{eq:E_inf_p_b}) and~(\ref{eq:e_inf_tot_b}), depicted in Fig.~\ref{fig:E_inf_b}
. The figure shows that the energy consumption per bit of developing a 3 layer - 10 node \ac{mlp} is $2.41\times 10^{-6}[J/b]$, which is approximately $3.7$ times higher than the energy consumption per bit of one inference ($E_{\mathrm{inf,p,b}}=6.57\times 10^{-7}[J/b]$). Furthermore, we can observe that the energy consumption per bit decreases with more inferences, dropping from $2.11\times 10^{-6} [J/b]$ for $100$ inferences to $7.73\times 10^{-7} [J/b]$ for $1000$ inferences, which is a $2.73$ times of improvement on energy efficiency. Moreover, we can observe that models with architectural blocks that require higher computational complexity need significantly more energy per bit during development compared to simpler models and their energy cost per bit of one inference. For instance, the transformer consumes $3.74 \times 10^{-4}~[\mathrm{J/b}]$, which means it is nearly $165$ and $20.7$ times less energy efficient per bit than the \ac{mlp} in making $100$ and $1000$ inferences, respectively. Fig.~\ref{fig:Ecal} further confirms that as the number of inferences \( \gamma \) in the current model increases, the \ac{ee} of the AI model lifecycle improves. 

\textit{Implementation guidelines:} \textcolor{black}{The open-source eCAL calculator (Python), can be used standalone to compute eCAL values using provided configuration files. Users can study custom scenarios, such as different transmit powers, overheads, or neural network configurations by editing these files. Thanks to its modular design, new protocols or neural architectures can be added via dedicated class  implementations. Python-based simulators can integrate it directly by importing its functionality, while other simulators may require minor additional development.} 
\section{Conclusions and Future Work}

\label{sec:conclusion}
In this work, we have proposed a novel metric, namely, eCAL. Unlike traditional metrics that focus only on energy required for transmission, computing infrastructure, or \ac{ai} models separately, eCAL can capture the overall energy cost of generating inferences in communication system during the entire lifecycle of a trained model. We have proposed a detailed methodology to determine the eCAL of an AI enhanced communication  system by breaking it down into various data manipulation components, such as data collection, preprocessing, training, evaluation, and inference, and analyzing the complexity and energy consumption of each component. The proposed metric demonstrates that the more a model is utilized, the more energy-efficient each inference becomes. For example, considering a simple MLP architecture, the energy consumption per bit for 100 inferences is $2.73$ times higher than for 1000 inferences. Based on the proposed methodology we also developed an open-source, modular eCAL calculator which enables fine-grained analysis of energy consumption across components, allowing developers to evaluate and optimize AI \ac{ee} in various deployment scenarios.

Through the proposed eCAL metric, this study lays the foundation for understanding the energy consumption of AI-enhanced emerging communication  architectures.
Future work may focus on extending this work beyond the assumptions, case studies, and configurations provided in this paper to cover existing and emerging communication system designs that are increasingly relying on \ac{ai} techniques for their automation and performance optimization, especially the emerging open radio access network (O-RAN) based cellular systems \textcolor{black}{and caching mechanism}.  The exploration of the relationship between model performance and eCAL for 6G verticals, related network slices, online, distributed, fine tuned and federated model lifecycles may also be worthy of pursuing. Finally, the development of dynamic energy management algorithms in view of a more sustainable operation is highly relevant. 


\paragraph*{Acknowledgements}
This work was supported in part by the HORIZON-MSCA-PF project TimeSmart (No. 101063721), the European Commission NANCY project (No. 101096456), and the Slovenian Research Agency under grants P2-0016, MN-0009-106, and J2-50071.


\balance

\bibliographystyle{./templates/IEEEtran}
\bibliography{IEEEabrv,bibliography}

\end{document}